\documentclass[runningheads]{llncs}

\usepackage{booktabs}   
\usepackage{subcaption} 

\usepackage{latexsym}
\usepackage{amssymb}
\usepackage{amsmath}
\usepackage{stmaryrd}
\usepackage[all]{xy}
\usepackage{graphicx}
\usepackage{hyperref}
\usepackage{multicol}
\usepackage{bm}
\usepackage{cancel}

\newcommand{\cons}{\!:\!}

\usepackage{tikz}
\usetikzlibrary{calc,arrows,shadows,positioning}

\newcommand{\upd}{\mathsf{add}}

\newcommand{\ignore}[1]{}

\newcommand{\arro}[1]{\xrightarrow{#1}} 
\newcommand{\boo}{\rightarrowtail} 
\newcommand{\hoo}{\hookrightarrow}

\newcommand{\comp}{\:|\:}

\renewcommand{\k}{\lambda}
\renewcommand{\k}{\ell}

\newcommand{\rlh}{\rightleftharpoons}
\newcommand{\lh}{\leftharpoondown}
\newcommand{\rh}{\rightharpoonup}

\usepackage{xcolor}

\newcommand{\red}[1]{{\color{red} #1}}

\newcommand{\blue}[1]{{\color{blue} #1}}

\newcommand{\id}{id}

\newcommand{\cC}{{\mathcal{C}}}
\newcommand{\cD}{{\mathcal{D}}}

\newcommand{\nil}{[\:]}

\def \tuple#1{\langle #1 \rangle}
\def \tuplef#1{( #1 )}
\def \tupleb#1{(\!( #1 )\!)}

\long\def\comment#1{}

\usepackage{accents}

\makeatletter
\providecommand{\leftsquigarrow}{%
  \mathrel{\mathpalette\reflect@squig\relax}%
}
\newcommand{\reflect@squig}[2]{%
  \reflectbox{$\m@th#1\rightsquigarrow$}%
}
\makeatother

\begin{document}

\title{%
From Reversible Computation to Checkpoint-Based
Rollback Recovery for Message-Passing Concurrent Programs%
\thanks{
This work has been partially supported by 
grant PID2019-104735RB-C41
funded by MCIN/AEI/ 10.13039/501100011033, by French ANR project DCore ANR-18-CE25-0007, and by  
\emph{Generalitat Valenciana} under grant CIPROM/2022/6 
(FassLow).
}
}

\titlerunning{From Reversible Computation to Checkpoint-Based
Rollback Recovery}

\author{Germ\'an Vidal}

\authorrunning{G. Vidal}

\institute{
  VRAIN, Universitat Polit\`ecnica de Val\`encia\\
\email{gvidal@dsic.upv.es}
}

\maketitle

\begin{abstract}
	The reliability of concurrent and distributed systems often 
	depends on some well-known techniques for fault tolerance. One 
	such technique is based on checkpointing and rollback recovery.  
	Checkpointing involves processes to take snapshots of their 
	current states regularly, so that a rollback recovery strategy
	is able to bring the system back to 
	a previous consistent state whenever a failure occurs. In this paper, 
	we consider a message-passing concurrent programming language 
	and propose a novel rollback recovery strategy that is based on 
	some explicit checkpointing operators and the use of 
	a (partially) reversible semantics for rolling back the system.\\[2ex]
  \emph{To appear in the Proceedings of the 
  19th International Conference on Formal Aspects of Component Software (FACS 2023).}	
\end{abstract}


\section{Introduction} \label{sec:intro}

The reliability of concurrent and distributed systems often depends 
on some	well-known techniques for fault tolerance. 
In this context, a popular approach is based on \emph{checkpointing}
and \emph{rollback recovery} (see, e.g., the survey by 
Elnozahy \emph{et al} \cite{EAWJ02}). 
Checkpointing requires
each process to take a snapshot of its state
at specific points in time. This state is stored in some stable
memory so that it can be accessed in case of a failure.
Then, if an unexpected error happens, a 
recovery strategy is responsible of rolling back the necessary processes
to a previous checkpoint so that we recover a consistent state of 
the complete system and normal execution can be safely resumed.

In this paper, we consider the definition of a rollback recovery strategy
based on three explicit operators: $\mathsf{check}$, 
$\mathsf{commit}$, and  $\mathsf{rollback}$. The first operator, 
$\mathsf{check}$, is used to define a checkpoint, thus saving the
current state of a process. The checkpoint is assigned a fresh
identifier $\tau$. Then, one can either \emph{commit} the
computation performed so far (up to the checkpoint), 
$\mathsf{commit}(\tau)$,
or roll back to the state immediately before the checkpoint,
$\mathsf{rollback}(\tau)$. 

There are several possible uses for these operators. 
For example, the functional and concurrent language Erlang
\cite{erlang} includes the usual try\_catch expression, which in
its simplest form is as follows:
``$
\mathtt{try}~e~\mathtt{catch}~\_:\_ \rightarrow e'~\mathtt{end}
$.''
Here, if the evaluation of expression $e$ terminates with some value, 
then try\_catch reduces to this value. Otherwise, if an exception is
raised (no matter the exception in this example 
since ``$\_:\_$'' catches all of them), 
the execution jumps to the catch statement and $e'$ is
evaluated instead. However, the actions performed during 
the incomplete evaluation of $e$ are not undone, 
which may give rise to
an inconsistent state of the system. Using the above operators,
we could write down a safer version of the 
try\_catch expression above as follows:\\[-1ex]

$
\mathtt{try}~\red{T = \mathsf{check}},~\red{X=\;} e,~\red{\mathsf{commit}(T)},~\red{X}~\mathtt{catch}~\_:\_ \rightarrow \red{\mathsf{rollback}(T)},~e'~\mathtt{end}
$\\[-1ex]

\noindent
In this case, we first introduce a checkpoint which reduces to
a fresh (unique) identifier, say $\tau$, and saves the current state
of the process as a side-effect; variable $T$ is bound to $\tau$.
Then, if the evaluation of expression $e$ completes successfully,
we gather the computed value in variable $X$, which
is returned after $\mathsf{commit}(\tau)$ removes the 
checkpoint.\footnote{Binding a temporal variable $X$ to the evaluation
of expression $e$ is required so that the try\_catch expression still
reduces to the same value of the original try\_catch expression; 
if we had just ``$T=\mathsf{check}, ~e,~\mathsf{commit}(T)$'' then this
sequence would reduce to the value returned by $\mathsf{commit}$
in Erlang, thus changing the original semantics.}
Otherwise, if an exception is raised, 
the execution jumps to the catch statement and
$\mathsf{rollback}(\tau)$ rolls back the process to the state
saved by checkpoint $\tau$ (possibly also rolling
back other processes in order to get a \emph{causally consistent} state; see below).

Our approach to rollback recovery is based 
on the notion of \emph{reversible computation} 
(see \cite{ACGKKKLMMNPPPUV20,GLMMPUV23}
 and references therein).
In principle, most programming languages are irreversible, 
in the sense that the execution of a statement cannot 
generally be undone. This is the case of Erlang, for instance.
Nevertheless, in these languages, one can still define a 
so-called \emph{Landauer embedding} \cite{Lan61} 
so that computations become reversible. 
Intuitively speaking, this operation amounts to
defining an instrumented semantics where states carry over
a \emph{history} with past states. 
While this approach may seem impractical
at first, there are several useful reversibilization techniques 
that are roughly based on this idea (typically including some 
optimization to reduce the amount of saved information, as in,
e.g., \cite{MHNHT07,NPV18}).

While the notion of reversible computation is quite natural 
in a sequential programming language, the extension to
concurrent and distributed languages presents some
additional challenges. Danos and Krivine \cite{DK04}
first introduced the notion of \emph{causal consistency}
in the context of a reversible calculus (Milner’s CCS \cite{Mil80}).
Essentially, in a causal consistent reversible setting,
\emph{an action cannot be undone
until all the actions that causally depend on this action
have been already undone}. E.g., we cannot undo
the spawning of a process until all the actions of this process
have been already undone; similarly, we cannot undo the sending
of a message until its reception (and the subsequent actions) have
been undone.
This notion of causality is closely
related with Lamport's ``happened before'' relation \cite{Lam78}.
In our work, we use a similar notion of causality to 
either propagate checkpoints and to perform causal consistent
rollbacks.

Our main contributions are the following. First, we propose the use
of three explicit operators for rollback recovery and provide their 
semantics. They can be used for defining a sort of \emph{safe}
transactions, as mentioned above, but not only. For instance, 
they could also be used as the basis of a reversible debugging
scheme where only some computations of interest 
are reversible, thus
reducing the run time overhead. Then, we define a rollback
semantics for the extended language that may proceed both
as the standard semantics (when no checkpoint
is active) or as a reversible semantics (otherwise). Finally, we
prove the soundness of our approach w.r.t.\ an 
\emph{uncontrolled} reversible semantics.

\section{A Message-Passing Concurrent Language} \label{sec:lang}

In this work, we consider a simple concurrent
language that mainly follows the \emph{actor model} \cite{HBS73}. 
Here, a running
system consists of a number of processes (or actors) that can 
(dynamically) create new processes and can only interact through 
message sending and receiving (i.e., no shared memory). 
This is the case, e.g., of (a significant subset of) 
the functional and concurrent
language Erlang \cite{erlang}.

In the following, we will ignore the sequential component of
the language (e.g., a typical eager functional programming language
in the case of Erlang) and will focus on its
concurrent actions:
\begin{itemize}
\item \emph{Process spawning}. A process may spawn new
processes dynamically. Each process is identified by a \emph{pid}
(a shorthand for \emph{p}rocess \emph{id}entifier), 
which is unique in a running system. 

\item \emph{Message sending}. A process can send a message
to any other process as long as it knows the pid of the target
process. This action is asynchronous.

\item  \emph{Message reception}. Messages need not  be
immediately consumed by the target process; rather, 
they are stored in an associated mailbox until they are
consumed (if any). We consider so-called
\emph{selective} receives, i.e., a process does not necessarily 
consume the messages in its mailbox in the same order they 
were delivered, since receive statements may impose additional 
constraints. When no message matches any constraint, the 
execution of the process is \emph{blocked} until a matching message 
reaches its mailbox.
\end{itemize}
In the following, we let $s,s',\ldots$ denote \emph{states}, typically 
including some environment, an expression (or statement) to be 
evaluated and, in some case, a stack. 
The structure of states is not relevant for the
purpose of this paper, though. 

A \emph{process} configuration is denoted
by a tuple of the form $\tuple{p,s}$, where $p$ is the pid of the process
and $s$ is its current state. 
Messages have the form
$(p,p',v)$ where $p$ is the pid of the sender, $p'$ that of the
receiver, and $v$ is the message value. 
A \emph{system}
is then denoted by a parallel composition of both processes and
(\emph{floating}) messages, as in \cite{LSZ19,LM20} (instead
of using a \emph{global mailbox}, as in \cite{LNPV18jlamp,LPV21}).
A floating message thus represents a message that has been
already sent but not yet delivered (i.e., the message is in the
network). Furthermore, in this work, process mailboxes
are abstracted away for simplicity, thus a floating message
can also represent a message that is actually
stored in a process mailbox.\footnote{In Erlang, the order of 
messages sent  directly from process $p$ to process $p'$ is 
preserved when they are  
all delivered; see~\cite[Section~10.8]{ErlangFAQ}.
We ignore this constraint 
for simplicity, but could be ensured by introducing
triples of the form $(p,p',vs)$ where $vs$ is a queue of messages
instead of a single message. }

Systems range over by $S$, $S'$, $S_1$, etc.
Here, the parallel composition operator is denoted by ``$|$'' and
considered commutative and associative. Therefore, two systems
are considered equal if they are the same up to associativity and
commutativity. 

As in \cite{LNPV18jlamp,LPV21},
the semantics of the language is defined in a modular way, so that the
labeled transition relations $\arro{}$ and $\boo$ model the evaluation
of expressions (or statements) and the evaluation of 
systems, respectively. 

\begin{figure}[t]
\centering
$
  \begin{array}{r@{~~}c}
    (\mathit{Seq}) & {\displaystyle
      \frac{s \arro{\mathsf{seq}} s' 
      }{\tuple{p,s} 
        \boo_{p,\mathsf{seq}} \tuple{p,s'}}
      }\\[3ex]

    (\mathit{Send}) & {\displaystyle
      \frac{s \arro{\mathsf{send}(p',v)} s' 
      }{\tuple{p,s} 
        \boo_{p,\mathsf{send}}
         (p,p',v) \comp \tuple{p,s'}}
      }\\[3ex]

      (\mathit{Receive}) & {\displaystyle
        \frac{s \arro{\mathsf{rec}(\kappa,cs)}
          s'~~\mbox{and}~~ \mathsf{matchrec}(cs,v) = cs_i}
          {(p',p,v) \comp
           \tuple{p,s} \boo_{p,\mathsf{rec}}
           \tuple{p,s'[\kappa\leftarrow cs_i])}}
      }\\[3ex]  

      (\mathit{Spawn}) & {\displaystyle
        \frac{s \arro{\mathsf{spawn}(\kappa,s_0)} 
          s'~~\mbox{and}~~ p'~\mbox{is a fresh pid}}
          {\tuple{p,s} \boo_{p,\mathsf{spawn}(p')}    
           \tuple{p,s'[\kappa\leftarrow p']}\comp \tuple{p',s_0}}
      }\\[3ex]

    (\mathit{Par}) & {\displaystyle
      \frac{S_1 \boo_e S'_1~~\mbox{and}~~\id(S'_1)\cap \id(S_2)=\emptyset}{S_1  \comp S_2 
         \boo_{e} S'_1 \comp S_2 
        }
      }      
  \end{array}
$
\caption{Standard semantics} \label{fig:standard-semantics}
\end{figure}

In the following, we skip the definition of the local semantics 
($\to$) since it is not necessary for our developments; 
we refer the interested reader to \cite{LNPV18jlamp,GV21tr}.
The rules of the operational semantics 
that define the reduction of systems 
is shown in Figure~\ref{fig:standard-semantics}.
The transition steps are labeled with the pid of the 
selected process and the considered action: 
$\mathsf{seq}$, $\mathsf{send}$, $\mathsf{rec}$,
or $\mathsf{spawn}(p')$, where $p'$ is the pid of the
spawned process.
Let us briefly explain these rules:
\begin{itemize}
\item Sequential, local steps are dealt with rule \emph{Seq}. Here,
we just propagate the reduction from the local level to the system
level.

\item Rule \emph{Send} applies when the local evaluation requires
sending a message as a side effect. The local step
$s \arro{\mathsf{send}(p',v)} s'$ is labeled with the information
that must flow from the local level to the system level:
the pid of the target
process, $p'$, and the message value, $v$. 
The system rule then adds a new message of the form $(p,p',v)$ 
to the system, where $p$ is the pid of the sender, $p'$ the pid of
the target process, and $v$ the message value.

\item In order to receive a message, the situation is somehow different.
Here, we need some information to flow both from the local level
to the system level (the clauses $cs$ of the receive statement)
and vice versa (the selected clause, $cs_i$, if any).
For this purpose, in rule \emph{Receive}, the label of the local step
includes a special variable $\kappa$ ---a sort of \emph{future}--- 
that denotes the position of the receive expression within state $s$. 
The rule then checks if there is
a floating message $v$ addressed to process $p$
that matches one of the constraints in $cs$. This is done by the
auxiliary function $\mathsf{matchrec}$, which returns the
selected clause $cs_i$ of the receive statement in case of a match
(the details are not relevant here).
Then, the reduction proceeds by binding $\kappa$ in $s'$ with the
selected clause $cs_i$, which we denote by $s'[\kappa\leftarrow cs_i]$. 

\item Rule \emph{Spawn} also requires a bidirectional flow of
information. Here, the label of the local step includes
the future $\kappa$ and the state of the new
process $s_0$. The rule then produces a fresh pid,
$p'$, adds the new process $\tuple{p',s_0}$ to
the system, and updates the state $s'$ by binding $\kappa$
to $p'$ (since $\mathsf{spawn}$ reduces to the pid of the new
process), which we denote by $s'[\kappa\leftarrow p']$.

\item Finally, rule \emph{Par} is  used to lift an evaluation
step to a larger system \cite{LSZ19}. The auxiliary function
$\id$ takes a system $S$ and returns the set of pids in $S$,
in order to ensure that new pids are indeed fresh in the complete
system.
\end{itemize}
In the following, $\boo^\ast$ denotes the 
transitive and reflexive closure of $\boo$. Given
systems $S_0,S_n$, we let $S_0 \boo^\ast S_n$ denote
a \emph{derivation} under the standard semantics.
When we want to consider the individual steps of a derivation,
we often write $S_0 \boo_{p_1,a_a} S_1 \boo_{p_2,a_2} 
\ldots \boo_{p_n,a_n} S_n$.
A reduction step usually consists of a number of applications
of rule $\mathit{Par}$ until a process, or a combination
of a process and a message, is selected, so that one of
the remaining rules can be applied ($\mathit{Seq}$,
$\mathit{Send}$, $\mathit{Receive}$ or $\mathit{Spawn}$).
We often omit the steps with rule 
$\mathit{Par}$ and only show the reductions on the selected 
process, i.e., $a_i \in\{\mathsf{seq},\mathsf{send},
\mathsf{rec},\mathsf{spawn}(p')\}$. 

An \emph{initial} system has the form $\tuple{p,s_0}$, 
i.e., it contains a single process. 
A system $S'$ is \emph{reachable} if there exists a derivation
$S \boo^\ast S'$ such that $S$ is an initial system. 
A derivation $S \boo^\ast S'$ is \emph{well-defined} under the
standard semantics if $S$ is a reachable system.

The semantics in Figure~\ref{fig:standard-semantics} 
applies to a significant subset of the 
programming language Erlang \cite{erlang},
as described, e.g., in \cite{GV21tr,LPV21}. 
However, for clarity, we will consider in the examples 
a much simpler notation which only shows some 
relevant information regarding the concurrent actions 
performed by each process. 
This is just a textual representation which makes
explicit process interaction but is not the actual program.
In particular, we describe the concurrent actions of a 
process by means of the following items:
\begin{itemize}
\item $p\!\leftarrow\! \mathsf{spawn}()$, for process spawning, 
where $p$ is the (fresh) pid returned by the call 
to $\mathsf{spawn}$ and assigned to the new process;
\item $\mathsf{send}(p,v)$, for sending a message, where
$p$ is the pid of the target process and $v$ the message value
(which could be a tuple including the process own pid in
order to get a reply, a common practice in Erlang);
\item $\mathsf{rec}(v)$, for receiving message $v$.
\end{itemize}
We will ignore sequential actions in the examples since they 
are not relevant for the purpose of this paper.	
	
\begin{figure}[t]
\centering
\begin{tikzpicture}
\draw[->,dashed,thick] (0,3.1) node[above]{$p_1$} -- (0,0);
\draw[->,dashed,thick] (3,3.1) node[above]{$p_2$} -- (3,0);
\draw[->,dashed,thick] (6,3.1) node[above]{$p_3$} -- (6,0);

\draw[->,dotted] (0,3.0) node[left]{$\mathsf{spawn}$} -- (3,2.9);

\draw[->,dotted] (3,2.7) node[left]{$\mathsf{spawn}$} -- (6,2.6);

\draw[->] (0,2.5) node[left]{$\mathsf{send}$} -- (3,2.4)
node[midway,above]{$v_1$} node[right]{$\mathsf{rec}$};

\draw[->] (3,2.2) node[left]{$\mathsf{send}$} -- (6,2.1)
node[midway,above]{$v_2$} node[right]{$\mathsf{rec}$};

\draw[->] (0,2.0) node[left]{$\mathsf{send}$} -- (3,1.9)
node[midway,above]{$v_3$} node[right]{$\mathsf{rec}$};

\draw[->] (6,1.8) node[right]{$\mathsf{send}$} -- (3,1.7)
node[midway,above]{$v_4$} node[left]{$\mathsf{rec}$};

\draw[->] (0,1.5) node[left]{$\mathsf{send}$} -- (3,1.4)
node[midway,above]{$v_5$} node[right]{$\mathsf{rec}$};

\draw[->] (3,1.1) node[left]{$\mathsf{send}$} -- (6,1.0)
node[midway,above]{$v_6$} node[right]{$\mathsf{rec}$};

\draw[->] (6,0.7) node[right]{$\mathsf{send}$} -- (3,0.6)
node[midway,above]{$v_7$} node[left]{$\mathsf{rec}$};
\end{tikzpicture}
\caption{Graphical representation of a reduction (time flows from top to bottom)}
\label{fig:ex-standard}
\end{figure}
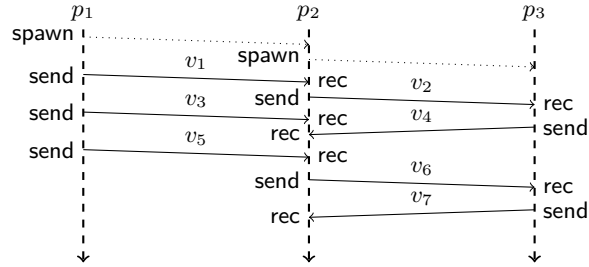

\begin{example} \label{ex1}
	Consider, for instance, a system with three processes with
	pids $p_1$ (the initial one), $p_2$, and $p_3$, which perform
	the following actions:\footnote{Note that, as mentioned before,
	we consider that processes can be dynamically spawned.
	Therefore, fresh pids must be sent to other processes
	in order to be known. This is abstracted away in our simple
	notation.}
	\[
	\begin{array}{l@{~~~~~~~~~~~~}l@{~~~~~~~~~~~~}l}
	 \mbox{\bf proc}~p_1 & \mbox{\bf proc}~p_2 & \mbox{\bf proc}~p_3 \\\hline
	 p_2\!\leftarrow\!\mathsf{spawn}()
	 & p_3\!\leftarrow\!\mathsf{spawn}()
	 & \mathsf{rec}(v_2) \\
	 \mathsf{send}(p_2,v_1)
	 & \mathsf{rec}(v_1)
	 & \mathsf{send}(p_2,v_4) \\
	 \mathsf{send}(p_2,v_3)
	 & \mathsf{send}(p_3,v_2)
	 & \mathsf{rec}(v_6) \\
	 \mathsf{send}(p_2,v_5) 
	 & \mathsf{rec}(v_3)
	 & \mathsf{send}(p_2,v_7) \\
	 & \mathsf{rec}(v_4) \\
	 & \mathsf{rec}(v_5) \\
	 & \mathsf{send}(p_3,v_6) \\
    & \mathsf{rec}(v_7) \\
	\end{array}
	\]
	A (partial) 
	derivation under the standard semantics 
	(representing a particular
	interleaving of the processes' actions) may proceed as follows,
	where we underline the selected process and message (if any) 
	at each step:\\[-1ex]
	
	$\hspace{-3ex}
	\begin{array}{llllllll}
	\underline{\tuple{p_1,s[p_2\!\leftarrow\!\mathsf{spawn}()]}}\\
	
	\boo_{p_1,\mathsf{spawn}(p_2)}
	{\tuple{p_1,s[\mathsf{send}(p_2,v_1)]}} \comp
	\underline{\tuple{p_2,s[p_3 \!\leftarrow\! \mathsf{spawn}()]}}\\
	
	\boo_{p_2,\mathsf{spawn}(p_3)}
	\underline{\tuple{p_1,s[\mathsf{send}(p_2,v_1)]}} \comp
	{\tuple{p_2,s[\mathsf{rev}(v_1)]}} \comp
	{\tuple{p_3,s[\mathsf{rec}(v_2)]}} \\

	\boo_{p_1,\mathsf{send}}
	{\tuple{p_1,s[\mathsf{send}(p_2,v_3)]}} \comp
	\underline{\tuple{p_2,s[\mathsf{rec}(v_1)]}} \comp
	{\tuple{p_3,s[\mathsf{rec}(v_2)]}} \comp
	\underline{(p_1,p_2,v_1)}\\

	\boo_{p_2,\mathsf{rec}}
	{\tuple{p_1,s[\mathsf{send}(p_2,v_3)]}} \comp
	\underline{\tuple{p_2,s[\mathsf{send}(p_3,v_2)]}} \comp
	{\tuple{p_3,s[\mathsf{rec}(v_2)]}}\\

	\boo_{p_2,\mathsf{send}}
	{\tuple{p_1,s[\mathsf{send}(p_2,v_3)]}} \comp
	{\tuple{p_2,s[\mathsf{rec}(v_3)]}} \comp
	\underline{\tuple{p_3,s[\mathsf{rec}(v_2)]}} \comp
	\underline{(p_2,p_3,v_2)}\\

	\boo_{p_3,\mathsf{rec}}
	\underline{\tuple{p_1,s[\mathsf{send}(p_2,v_3)]}} \comp
	{\tuple{p_2,s[\mathsf{rec}(v_3)]}} \comp
	{\tuple{p_3,s[\mathsf{send}(p_2,v_4)]}} \\

	\boo_{p_1,\mathsf{send}}
	{\tuple{p_1,s[\mathsf{send}(p_2,v_5)]}} \comp
	\underline{\tuple{p_2,s[\mathsf{rec}(v_3)]}} \comp
	{\tuple{p_3,s[\mathsf{send}(p_2,v_4)]}} \comp
	\underline{(p_1,p_2,v_3)} \\

	\boo \ldots
%
%
%
%
%
%
%
%
	\end{array}
	$\\

	\noindent
   where a state of the form $s[op]$ denotes an arbitrary state 
   where the next operation to be reduced is $op$. We omit some
   intermediate steps which are not relevant here.
	A graphical representation of the complete reduction
	can be found in Fig.~\ref{fig:ex-standard}. 
\end{example}
We note that programs can exhibit an iterative behavior
through recursive function calls.\footnote{For instance, 
a typical server process is defined by a function that 
waits for a client request, process it, and then makes 
a recursive call, possibly with a modified \emph{state}; 
this is a common pattern in Erlang.}
This is hidden in our examples since we do not show 
explicitly sequential operations.

\section{Checkpoint-Based Rollback Recovery}

In this section, we present our approach to rollback recovery
in a message-passing concurrent language. Essentially, our
approach is based on defining an instrumented semantics
with two modes: a ``normal'' mode, which proceeds 
similarly to the standard semantics, and a 
``reversible'' mode, where actions can be undone and, thus, 
can be used for rollbacks.

\subsection{Basic Operators}

We consider three explicit operators to
control rollback recovery: $\mathsf{check}$,
$\mathsf{commit}$, and $\mathsf{rollback}$.
%
Intuitively speaking, they 
proceed as follows:
\begin{itemize}
\item $\mathsf{check}$ introduces a \emph{checkpoint} for the 
current process. The reduction of \textsf{check}
returns a fresh identifier, $\tau$, associated to the checkpoint; note
that nested checkpoints are possible.

\item $\mathsf{commit}(\tau)$ can then 
be used to discard the state saved in checkpoint $\tau$. 
In our context, $\mathsf{check}$ implies turning the reversible
mode on and $\mathsf{commit}$ turning it off (when no
more active checkpoints exist).

\item Finally, $\mathsf{rollback}(\tau)$ starts a backward
computation, undoing all the actions of the process (and their
causal dependencies from other processes) up to  the call to 
$\mathsf{check}$ (including) that introduced $\tau$.
\end{itemize}

\begin{figure}[t]
\centering
$
  \begin{array}{c}
    (\mathit{Check}) ~ {\displaystyle
      \frac{} 
      {\theta,\mathsf{check} \arro{\mathsf{check}(\kappa)}
      \theta,\kappa}
      }\\[5ex]

    (\mathit{Commit}) ~ {\displaystyle
      \frac{}{\theta,\mathsf{commit}(\tau) \arro{\mathsf{commit(\tau)}}
      \theta,ok}
      } 
\hspace{3ex}
    (\mathit{Rollback}) ~ {\displaystyle
      \frac{}{\theta,\mathsf{rollback}(\tau) \arro{\mathsf{rollback}(\tau)}
      \theta,ok}
      }\\[2ex]
  \end{array}
$
\caption{Rollback recovery operators} \label{fig:operators}
\end{figure}

\noindent
The local reduction rules for the new operators are very simple and
can be found in Figure~\ref{fig:operators}. 
Here, we consider that a local state
consists of an environment (a variable substitution) and an
expression (to be evaluated), but it could be straightforwardly
extended to other state configurations (e.g., configurations that also  include a stack, as in \cite{GV21}). 

Rule \emph{Check} reduces the call to a future, $\kappa$, which
also occurs in the label of the transition step. 
As we will see in the next section, 
the corresponding rule in the system semantics will perform
the associated side-effect (creating a checkpoint) and will also
bind $\kappa$ with the (fresh) identifier for this checkpoint.		
Rules \emph{Commit} and \emph{Rollback} just pass the 
corresponding information to the system semantics in order
to do the associated side effects. Both rules reduce the call
to the constant ``$ok$'' (an atom used in Erlang when a
function call does not return any value).

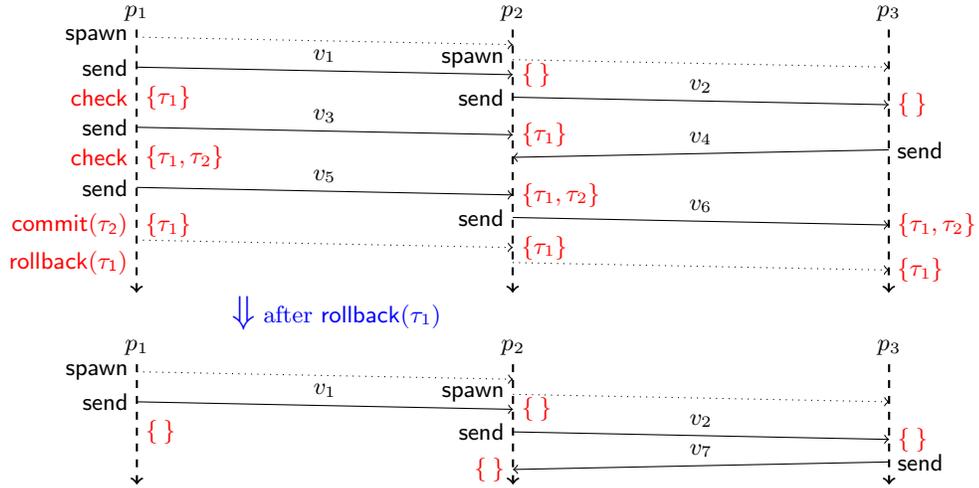
\begin{figure}[t]
\centering
\begin{tikzpicture}
\draw[->,dashed,thick] (0,3.5) node[above]{$p_1$} -- (0,0);
\draw[->,dashed,thick] (5,3.5) node[above]{$p_2$} -- (5,0);
\draw[->,dashed,thick] (10,3.5) node[above]{$p_3$} -- (10,0);

\draw[->,dotted] (0,3.4) node[left]{$\mathsf{spawn}$} -- (5,3.3);

\draw[->,dotted] (5,3.1) node[left]{$\mathsf{spawn}$} -- (10,3.0);

\draw[->] (0,3.0) node[left]{$\mathsf{send}$} -- (5,2.9)
node[midway,above]{$v_1$} node[right]{\red{$\{\:\}$}};

\draw[->] (5,2.6) node[left]{$\mathsf{send}$} -- (10,2.5)
node[midway,above]{$v_2$} node[right]{\red{$\{\:\}$}};

\draw (0,2.6) node[left]{\red{$\mathsf{check}$}}
node[right]{\red{$\{\tau_1\}$}};

\draw[->] (0,2.2) node[left]{$\mathsf{send}$} -- (5,2.1)
node[midway,above]{$v_3$} node[right]{\red{$\{\tau_1\}$}};

\draw[->] (10,1.9) node[right]{$\mathsf{send}$} -- (5,1.8)
node[midway,above]{$v_4$}; 

\draw (0,1.8) node[left]{\red{$\mathsf{check}$}}
node[right]{\red{$\{\tau_1,\tau_2\}$}};

\draw[->] (0,1.4) node[left]{$\mathsf{send}$} -- (5,1.3)
node[midway,above]{$v_5$} node[right]{\red{$\{\tau_1,\tau_2\}$}};

\draw[->] (5,1.0) node[left]{$\mathsf{send}$} -- (10,0.9)
node[midway,above]{$v_6$} node[right]{\red{$\{\tau_1,\tau_2\}$}};

\draw (0,0.9) node[left]{\red{$\mathsf{commit(\tau_2)}$}}
node[right]{\red{$\{\tau_1\}$}};

\draw[->, dotted] (0,0.7) -- (5,0.6) node[right]{\red{$\{\tau_1\}$}};

\draw[->, dotted] (5,0.4) -- (10,0.3) node[right]{\red{$\{\tau_1\}$}};

\draw (0,0.4) node[left]{\red{$\mathsf{rollback(\tau_1)}$}};

\end{tikzpicture}

\mbox{}\hspace{-24ex}{\blue{{\Large $\Downarrow$} after $\mathsf{rollback}(\tau_1)$}}

\begin{tikzpicture}
\draw[->,dashed,thick] (0,1.6) node[above]{$p_1$} -- (0,0);
\draw[->,dashed,thick] (5,1.6) node[above]{$p_2$} -- (5,0);
\draw[->,dashed,thick] (10,1.6) node[above]{$p_3$} -- (10,0);

\draw[->,dotted] (0,1.5) node[left]{$~~~~~~~\mathsf{spawn}$} -- (5,1.4);

\draw[->,dotted] (5,1.2) node[left]{$\mathsf{spawn}$} -- (10,1.1);

\draw[->] (0,1.1) node[left]{$\mathsf{send}$} -- (5,1.0)
node[midway,above]{$v_1$} node[right]{\red{$\{\:\}$}};

\draw (0,0.7) node[right]{\red{$\{\:\}$}};

\draw[->] (5,0.7) node[left]{$\mathsf{send}$} -- (10,0.6)
node[midway,above]{$v_2$} node[right]{\red{$\{\:\}$}};

\draw[->] (10,0.3) node[right]{$\mathsf{send}$} -- (5,0.2)
node[midway,above]{$v_7$} node[left]{\red{$\{\:\}$}};
\end{tikzpicture}

\caption{Graphical representation of a reduction with 
$\mathsf{check}$, $\mathsf{commit}$, and $\mathsf{rollback}$}
\label{fig:ex2}
\end{figure}

\begin{example} \label{ex2}
Consider again a program that performs the concurrent actions
of Example~\ref{ex1}, where we now 
add a couple of checkpoints, a commit, and a rollback to process 
$p_1$. Here, 
we let $\tau\!\leftarrow\!\mathsf{check}$ denote that $\tau$
is the (fresh) identifier returned by the call to $\mathsf{check}$:
	\[
	\begin{array}{l@{~~~~~~~~~~~~}l@{~~~~~~~~~~~~}l}
	 \mbox{\bf proc}~p_1 & \mbox{\bf proc}~p_2 & \mbox{\bf proc}~p_3 \\\hline
	 p_2\!\leftarrow\!\mathsf{spawn}()
	 & p_3\!\leftarrow\!\mathsf{spawn}()
	 & \mathsf{rec}(v_2) \\
	 
	 \mathsf{send}(p_2,v_1)
	 & \mathsf{rec}(v_1)
	 & \mathsf{send}(p_2,v_4) \\

	 \red{\tau_1\!\leftarrow\!\mathsf{check}}
	 & \mathsf{send}(p_3,v_2)
	 & \mathsf{rec}(v_6) \\
	 
	 \mathsf{send}(p_2,v_3) 
	 & \mathsf{rec}(v_3)
	 & \mathsf{send}(p_2,v_7) \\

     \red{\tau_2\!\leftarrow\!\mathsf{check}}
	 & \mathsf{rec}(v_4) \\

     \mathsf{send}(p_2,v_5)
	 & \mathsf{rec}(v_5) \\

	 \red{\mathsf{commit}(\tau_2)}
	 & \mathsf{send}(p_3,v_6) \\

    \red{\mathsf{rollback}(\tau_1)}
    & \mathsf{rec}(v_7) \\
	\end{array}
	\]
A graphical representation of the new execution
can be found in Fig.~\ref{fig:ex2}. Intuitively speaking,
it proceeds as follows:
\begin{itemize}
\item Process $p_1$ calls function $\mathsf{check}$, which 
creates a checkpoint with identifier $\tau_1$. This checkpoint
is propagated to $p_2$ when sending message $v_3$, so
$p_2$ turns the reversible mode on too. 
\item Then, $p_1$ creates 
another checkpoint, $\tau_2$, so we have two active
checkpoints. These checkpoints are propagated to $p_2$
by message $v_5$ and also to $p_3$ by
message $v_6$. At this point, all three
processes have the reversible mode on.

\item Now, $p_1$ calls $\mathsf{commit}(\tau_2)$, 
so checkpoint $\tau_2$ is not active anymore in $p_1$. 
This is also propagated to both $p_1$ and $p_2$.
Nevertheless, the reversible mode is still on in all three 
processes since $\tau_1$ is still alive.

\item Then, $p_1$ calls $\mathsf{rollback}(\tau_1)$ so
$p_1$ undoes all its actions up to (and including) the first
call to $\mathsf{check}$. For this rollback to be causal
consistent, $p_2$ rolls back to the point immediately
before receiving message $v_3$, and $p_3$ to the 
point immediately before receiving message $v_6$.

\item Finally, all three processes are back to normal,
irreversible mode (no checkpoint is active), and $p_3$
sends message $v_7$ to $p_2$. The actions in the second 
diagram are all irreversible.
\end{itemize}
\end{example}

\subsection{A Reversible Semantics for Rollback Recovery}

Let us now consider the design of a \emph{reversible}
semantics for rollback recovery.
Essentially, the operators can be modeled as follows:
\begin{itemize}
\item The reduction of $\mathsf{check}$ creates a
checkpoint, which turns on the reversible
mode of a process as a side-effect 
(assuming it was not already on). As in \cite{LNPV18jlamp,LPV21}, 
reversibility is achieved by defining an appropriate 
\emph{Landauer embedding} \cite{Lan61}, 
i.e., by adding a history of the computation to each process 
configuration.\footnote{For clarity of exposition, the
complete state is saved at each state. Nevertheless, 
an optimized history could also be defined; 
see, e.g., \cite{MHNHT07,NPV18}. } 
A checkpoint is propagated to other processes
when a causally dependent action is performed 
(i.e., $\mathsf{spawn}$ and $\mathsf{send}$);
following the terminology of \cite{EAWJ02},
these checkpoints are called
\emph{forced} checkpoints.

\item A call of the form 
$\mathsf{commit}(\tau)$ removes $\tau$ from
the list of \emph{active} checkpoints of a process,
turning the reversible mode off when the list of
active checkpoints is empty.
Forced checkpoints in other processes with
the same identifier $\tau$ (if any) are also removed from
the corresponding sets of active checkpoints.
\item Finally, the reduction of $\mathsf{rollback}(\tau)$ involves 
undoing all the steps of a given process up to the checkpoint $\tau$
\emph{in a causal consistent way}, i.e., possibly also undoing causally 
dependent actions from other processes.
\end{itemize}
Unfortunately, this apparently simple model presents a 
problem. Consider, for instance, a
process $p$ that performs the following actions:
\begin{equation}
\hspace{-2ex}  \label{eqn:problem}
p:=\{
\rlap{$\overbrace{\phantom{\mathsf{check}(\tau_1),
\mathsf{send}(p',v),\mathsf{check}(\tau_2),\ldots,
\mathsf{commit}(\tau_1)}}$}
\mathsf{check}(\tau_1),\mathsf{send}(p',v),
\underbrace{\mathsf{check}(\tau_2),\ldots,
\mathsf{commit}(\tau_1),\ldots,
\mathsf{rollback}(\tau_2)},\ldots\} \hspace{2ex} \tag{*}
\end{equation}
Here, we can observe that the pairs 
$\mathsf{check}$-$\mathsf{commit}$ 
and $\mathsf{check}$-$\mathsf{rollback}$
are not well balanced. As a consequence, we commit
checkpoint $\tau_1$ despite the fact that a rollback like
$\mathsf{rollback}(\tau_2)$ may
bring the computation back to the point immediately before
$\mathsf{check}(\tau_2)$, where $\tau_1$ should be alive,
thus producing an inconsistent state.
We could recover $\tau_1$ when undoing the call 
$\mathsf{commit}(\tau_1)$. However, 
this is not only a \emph{local} 
problem, since $\mathsf{commit}(\tau_1)$ may have 
also removed the checkpoint from other processes 
($p'$, in the example).

We could solve the above problem by considering that a call
to $\mathsf{commit}$ may introduce new causal dependencies.
Intuitively speaking, one could treat the propagation of 
$\mathsf{commit}$ to other process as the sending of a 
message. Therefore, a causal consistent rollback would often
require undoing all subsequent actions in these processes 
before undoing the call to $\mathsf{commit}$. Unfortunately,
this solution would require all processes to keep the
reversible mode on all the time, which is precisely what
we want to avoid.

A very simple workaround comes immediately to mind:
require the programmer to write well-balanced
pairs $\mathsf{check}$-$\mathsf{commit}$ 
and $\mathsf{check}$-$\mathsf{rollback}$.
In this case, a rollback that undo a call to $\mathsf{commit}$ will
also undo the corresponding call to $\mathsf{check}$,
so the inconsistent situation above would no longer be possible.
However, this solution is not acceptable when forced
checkpoints come into play. For instance, 
in example (\ref{eqn:problem}) above,
if we replace $\mathsf{check}(\tau_2)$  by a receive operation from
some process where checkpoint $\tau_2$ is active, 
the same inconsistent
state can be reproduced. In this case, though, we cannot
require the programmer to avoid such situations since
they are unpredictable.

For all the above, in this work we do not impose any constraint
on the use of the new operators but
propose the following solution:
When a call of the form $\mathsf{commit}(\tau)$ occurs, 
we check in the process' history whether 
$\tau$ is the \emph{last} active checkpoint
of the process (either proper or forced). 
If this is the case, the commit is executed.
Otherwise, it is \emph{delayed} until the condition is met.

Our configurations will now have three 
additional fields:
the set of active checkpoints, the set of delayed commits,
and a history (for reversibility):

\begin{definition}[rollback configuration]
	A \emph{forward} configuration is defined as a tuple 
	$\tuplef{\cC,\cD,h,\tuple{p,s}}$ where $\cC$ is a set of 
	(active) checkpoint identifiers, $\cD$ is a set of 
	delayed commits, $h$ is a history, and
	$\tuple{p,s}$ is a pid and a (local) state, similarly to 
	the standard semantics.
	A \emph{backward} configuration has the form 
	$\tupleb{\cC,\cD,h,\tuple{p,s}}^\tau$ where $\tau$
	is the target of a rollback request. 

	A (rollback) \emph{configuration} is either a 
	forward or a backward configuration.
\end{definition}
As for the \emph{messages}, they now have the form 
$(\red{\cC},p,p',\{\red{\ell},v\})$. 
Here, we can distinguish two differences
w.r.t.\ the standard semantics: first, we add a set of 
active checkpoints, $\cC$, which should be propagated to
the receiver as \emph{forced} checkpoints; secondly, 
the message value is wrapped with a tag to uniquely identify it
(as in \cite{LNPV18jlamp}, in order to 
distinguish messages with the same value).

As in the standard semantics, a \emph{system} is then a parallel
composition of (rollback) configurations and floating messages. 

In the following, we let $\nil$ denote an empty list 
and $x\cons xs$ a list with head $x$ and tail $xs$. 
A process \emph{history} $h$ is then represented by a list
of the following elements: 
$\mathsf{seq}$, $\mathsf{send}$, $\mathsf{rec}$, 
$\mathsf{spawn}$,  $\mathsf{check}$,
and $\mathsf{commit}$.
Each one is denoted by a term 
containing enough information to undo the 
corresponding reduction step,\footnote{The reader can
compare the rules in Fig.~\ref{fig:rollback-semantics-forward}
and their inverse counterpart in 
Fig.~\ref{fig:rollback-semantics-backward}.} 
except for the case of 
$\mathsf{commit}$, whose side-effects are irreversible,
as argued above. To be more precise,
\begin{itemize}
\item all terms store the current state ($s$).
\item for sending a message, the corresponding term also stores 
the pid of the target process ($p'$) and the message 
tag ($\ell$); 
\item for receiving a message, the term also stores
two sets of active checkpoints (the active ones and those
received from a message as forced checkpoints), the pid of
the sender ($p'$), the message tag ($\ell$), and
the message value ($v$);
\item for spawning a process, the corresponding term also
includes the fresh pid of the new process ($p'$);
\item and, finally, for $\mathsf{check}$ and $\mathsf{commit}$, 
it also stores the checkpoint identifier ($\tau$).
\end{itemize}
A delayed commit is represented as a triple
$\tuple{\tau,h,P}$, where $\tau$ is a checkpoint identifier,
$h$ is a history, and
$P$ is a set of pids (the pids of the processes where a
forced checkpoint $\tau$ has been propagated).

\subsubsection{Forward Rules.}

\begin{figure}[t]
\centering
$\hspace{-2ex}
  \begin{array}{r@{~~}c}
      (\underline{\mathit{Seq}}) & {\displaystyle
      \frac{s \arro{\mathsf{seq}} s'
      ~\mbox{and}~\red{\upd_\cC(\mathsf{seq}(s),h) = h'}
      }{\tuplef{\cC,\cD,h,\tuple{p,s}} 
        \hoo_{p,\underline{\mathsf{seq}}} 
        \tuplef{\cC,\cD,\red{h'},\tuple{p,s'}}}
      }\\[3ex]

    (\underline{\mathit{Send}}) & {\displaystyle
      \frac{s \arro{\mathsf{send}(p',v)} s', 
      ~\red{\ell~\mbox{is a fresh symbol}},~\mbox{and}~
      \red{\upd_\cC(\mathsf{send}(s,p',\ell),h) = h'}
      }{\tuplef{\cC,\cD,h,\tuple{p,s}} 
        \hoo_{p,\underline{\mathsf{send}}(\red{\ell})}
         (\red{\cC},p,p',\{\red{\ell},v\}) \comp 
         \tuplef{\cC,\cD,\red{h'},\tuple{p,s'}}}
      }\\[3ex]
      
      (\underline{\mathit{Receive}}) &  {\displaystyle
        \frac{s \arro{\mathsf{rec}(\kappa,cs)}
          s',~ \mathsf{matchrec}(cs,v) = cs_i,~\mbox{and}~
         \red{\upd_\cC(\mathsf{rec}(\cC'\setminus \cC,\cC',s,p',\ell,v),h) = h'}
         }
          {(\red{\cC'},p',p,\{\red{\ell},v\}) \comp
           \tuplef{\red{\cC},\cD,h,\tuple{p,s}} 
           \hoo_{p,\underline{\mathsf{rec}}(\red{\ell})}
           \tuplef{\red{\cC\cup\cC'},\cD,\red{h'},\tuple{p,s'[\kappa\leftarrow cs_i]}}}
      }\\[3ex]  
      
   (\underline{\mathit{Spawn}}) & {\displaystyle
        \frac{s \arro{\mathsf{spawn}(\kappa,s_0)} 
          s',~ p'~\mbox{is a fresh pid},~\mbox{and}~
          \red{\upd_\cC(\mathsf{spawn}(s,p'),h) = h'}
          }
          {\tuplef{\red{\cC},\cD,h,\tuple{p,s}} \hoo_{p,\underline{\mathsf{spawn}}(p')}    
           \tuplef{\cC,\cD,\red{h'},\tuple{p,s'[\kappa\leftarrow p']}}\comp \tuplef{\red{\cC},\nil,\tuple{p',s_0}}}
      }\\[3ex]

    (\underline{\mathit{Check}}) & {\displaystyle
        \frac{s \arro{\mathsf{check}(\kappa)} s'~\mbox{and}~
        \red{\tau~\mbox{is a fresh identifier}}
      }  
          {\tuplef{\cC,\cD,h,\tuple{p,s}} \hoo_{p,\underline{\mathsf{check}}(\tau)}
           \tuplef{\red{\cC\cup\{\tau\}},\cD,\red{\mathsf{check}(\tau,s)}\cons h,\tuple{p,s'[\kappa\leftarrow \tau])}}}
      }\\[3ex]  
      
  (\underline{\mathit{Commit}}) & {\displaystyle
      \frac{s \arro{\mathsf{commit}(\tau)} s',~ 
      \red{\mathit{last}_\tau(h)=\mathit{true}},~
      dp_\tau(h)=P,~\mbox{and}~
	  \red{\mathit{propagate}(\tau,P)}
	  }
      {\tuplef{\cC,\cD,h,\tuple{p,s}} 
         \hoo_{p,\underline{\mathsf{commit}}(\tau)} 
        \tuplef{\cC\setminus\{\red{\tau}\},\cD,\red{\mathsf{commit}(\tau,s)}\cons h,\tuple{p,s'}}
        }
      }\\[3ex]
      
   & {\displaystyle
      \frac{s \arro{\mathsf{commit}(\tau)} s',~ 
      \red{\mathit{last}_\tau(h)=\mathit{false}},~\mbox{and}~
      dp_\tau(h)=P
	  }
      {\tuplef{\cC,\cD,h,\tuple{p,s}} 
         \hoo_{p,\underline{\mathsf{delay}}(\tau)} 
        \tuplef{\cC\setminus\{\red{\tau}\},\cD\cup\{\red{\tuple{\tau,h,P}}\},\red{\mathsf{commit}(\tau,s)}\cons h,\tuple{p,s'}}
        }
      }\\[3ex]
      
   (\underline{\mathit{Delay}}) & {\displaystyle
      \frac{{\mathit{last}_\red{\tau}(\red{h'})=\mathit{true}}
      ~\mbox{and}~
	  \red{\mathit{propagate}(\tau,P)}
	  }
      {\tuplef{\red{\cC},\cD\cup\{\tuple{\red{\tau},\red{h'},P}\},h,\tuple{p,s}}
         \hoo 
        \tuplef{\cC,\cD,h,\tuple{p,s}}
        }
      }\\[3ex]      
            
    (\underline{\mathit{Rollback}}) & {\displaystyle
      \frac{s \arro{\mathsf{rollback}(\tau)} s' 
      }{\tuplef{\cC,\cD,h,\tuple{p,s}} 
         \hoo_{\red{p,\underline{\mathsf{rollback}}(\tau)}} 
        \tupleb{\cC,\cD,h,\tuple{p,s'}}^\tau
        }
      }\\[3ex]
      
      (\underline{\mathit{Par}}) & {\displaystyle
      \frac{S_1 \hoo_\red{l} S'_1~~\mbox{and}~~\id(S'_1)\cap \id(S_2)=\emptyset}{S_1  \comp S_2 
         \hoo_{\red{l}} S'_1 \comp S_2 
        }
      }      

  \end{array}
  $
\caption{Rollback recovery semantics: forward rules} 
\label{fig:rollback-semantics-forward}
\end{figure}

The \emph{forward} reduction rules of the rollback semantics 
are shown in Figure~\ref{fig:rollback-semantics-forward}. 
The main difference with the standard semantics is that, 
now, some process configurations
include a \emph{history} with enough information to undo any
reduction step. Here, we follow the same strategy as in \cite{LNPV18jlamp,LPV21} in order to define
a scheme for reversible debugging.\footnote{In contrast to \cite{LNPV18jlamp,LPV21}, however, we do not
need to undo every possible step, but only those steps that are
performed when there is at least one active checkpoint.
This is why we added $\cC$: to store the set 
of \emph{active} checkpoints
(i.e., checkpoints without a corresponding commit/rollback yet).
Observe that we might have several active checkpoints not only 
because nested checkpoints are possible, 
but because of \emph{forced}
checkpoints propagated by process spawning and message
sending.}

Essentially, the first four rules of the semantics 
can behave either as the
standard semantics or as a \emph{reversible} semantics, 
depending on whether $\cC$ is
empty or not. For conciseness, we avoid duplicating all rules by
introducing the auxiliary function $\upd$ to update the history
only when there are active checkpoints:
$\upd_\cC(a,h)= h$
if $\cC=\emptyset$, and $\upd_{\cC}(a,h)=a\cons h$ 
otherwise.

As mentioned above, the reversible mode 
is propagated through message sending
and receiving. This is why messages now include the set of 
active checkpoints $\cC$. As can be seen in 
rule \underline{\emph{Receive}}, the process receiving 
the message updates
its active checkpoints with those in the message. This is necessary
for rollbacks to be causally consistent. Note that, in the associated
term in the history, $\mathsf{rec}(\cC'\setminus\cC,\cC',s,p',\ell,v)$,
$\cC'\setminus \cC$ denotes the \emph{forced} checkpoints 
introduced by the received message. 

Similarly, the reversible mode is also propagated by 
process spawning: rule \underline{\emph{Spawn}} adds the
current set of active checkpoints $\cC$ (which might be empty)
to the new process.

As for the new rules, \underline{\emph{Check}} 
produces a fresh identifier,
$\tau$, and binds the future, $\kappa$, to this identifier.
Moreover, it also adds $\tau$ to the current set of active checkpoints.
In particular, if $\cC$ is empty, this step turns
the reversible mode on.

\underline{\emph{Commit}} includes two transition rules,
depending on whether the commit can be done or it should be 
delayed. We use the auxiliary Boolean function
$\mathit{last}$ so that $\mathit{last}_\tau(h)$
checks whether $\tau$ is the last checkpoint 
of the process according to history $h$, i.e., whether
the last $\mathsf{check}$ or $\mathsf{rec}$ term in $h$ 
has either the form $\mathsf{check}(\tau,s')$
or $\mathsf{rec}(\cC',\ldots)$ with $\tau\in\cC'$.
Note that we do
not need to consider forced checkpoints introduced by 
process spawning since they cannot occur after a
call to $\mathsf{check}$ (they are always introduced when
spawning the process).
Then, if the call to function $\mathit{last}$ returns \emph{true}, 
we remove the
checkpoint identifier from $\cC$ and from all processes where 
this checkpoint was propagated (as a forced checkpoint). 
Here, the auxiliary funcion $dp_\tau$ takes a history
and returns all pids which have causal
dependencies with the current process according to $h$, i.e.,
\begin{itemize}
\item $\red{p} \in dp_\tau(h)$ if
$\mathsf{spawn}(s,\red{p})$ occurs in $h$;
\item $\red{p} \in dp_\tau(h)$
if $\mathsf{send}(s,\red{p},\ell)$ occurs in $h$.
\end{itemize}
Now, we want to propagate the effect of commit to all processes
in $dp_\tau(h)$ in order to remove $\tau$ from their set of
active checkpoints.
One could formalize this process with a few more transition
rules and a new kind of configuration. 
However, for simplicity,
we represent it by means of an 
auxiliary function $\mathit{propagate}$ so that 
$\mathit{propagate}(\tau,P)$ always returns $\mathit{true}$
and performs the following side-effects:
\begin{enumerate}
\item for each $p'\in P$, we look for the process
with pid $p'$, say $\tuplef{\cC',\cD',h',\tuple{p',s'}}$;
\item if $\tau\not\in\cC'$ (it is not an active checkpoint
of process $p'$), we are done;
\item otherwise ($\tau\in\cC')$, we remove $\tau$ from $\cC'$
and repeat the process, i.e., we compute $P'=dp_\tau(h')$
and call $propagate(\tau,P')$.
\end{enumerate}
Termination is ensured since the number of processes is
finite and a process where 
$\tau$ is not active will eventually be reached.
In practice, commit propagation can be implemented 
by sending (asynchronous) messages to the involved
processes.
Note that the semantics would be sound even if 
commit operations were not propagated, so doing it is
essentially a matter of efficiency (a sort of 
\emph{garbage collection} to avoid recording
actions that are not really necessary).

On the other hand, if the call to function $\mathit{last}$ 
returns \emph{false}, the checkpoint is moved from 
$\cC$ to $\cD$ as a delayed commit (second rule of
$\underline{\mathit{Commit}}$).
Eventually, rule \underline{\emph{Delay}} becomes applicable
and proceeds similarly to the first rule of
\underline{\emph{Commit}} but considering the delayed commit.
We do not formalize a particular strategy for firing rule
$\underline{\mathit{Delay}}$, but a simple strategy would only
fire this rule only 
when some checkpoint is removed from
the set of active checkpoints of a process.

Rule \underline{\emph{Rollback}} simply changes the 
forward configuration to a backward configuration, also adding the
superscript $\tau$ to \emph{drive} the rollback. 
Therefore, the forward rules are no longer
applicable to this process (and the backward rules in 
Figure~\ref{fig:rollback-semantics-backward} can be applied
instead).

Finally, rule \underline{\emph{Par}} is identical to that in the standard
semantics. The only difference is that, now, function $\id(S)$ 
returns the set of pids, message tags, and checkpoints in $S$.

\subsubsection{Backward Rules.}

\begin{figure}[t]
  \centering
  $\hspace{-2ex}
  \begin{array}{r@{~~}c}
    (\overline{\mathit{Seq}}) & {\displaystyle
      \tupleb{\cC,\cD,\mathsf{seq}(\red{s})\cons h,\tuple{p,s'}}^\tau
      \hoo_{p,\overline{\mathsf{seq}}} 
      \tupleb{\cC,\cD,h,\tuple{p,\red{s}}}^\tau
      }
      \\[2ex]

    (\overline{\mathit{Send}}) & {\displaystyle
      (\red{\cC},p,p',\{\red{\ell},v\}) \comp 
         \tupleb{\red{\cC},\cD,\mathsf{send}(\red{s},p',\red{\ell})\cons h,
         \tuple{p,s'}}^\tau
        \hoo_{p,\overline{\mathsf{send}}(\red{\ell})}         
         \tupleb{\red{\cC},\cD,h,\tuple{p,\red{s}}}^\tau          
      }\\[2ex]

 & {\displaystyle
    \begin{array}[t]{l}
         \tupleb{{\cC},\cD,\mathsf{send}({s},\red{p'},{\ell})\cons h,\tuple{p,s'}}^\tau
         \comp \tuplef{\cC',\cD',h',\tuple{\red{p'},s''}}
         \\
        \hspace{15ex}\hoo
         \tupleb{{\cC},\cD,\mathsf{send}({s},{p'},{\ell})\cons h,\tuple{p,s'}}^\tau
         \comp \tupleb{\cC',\cD',h',\tuple{{p'},s''}}^\tau
      \end{array}
      }\\[5ex]
      
      (\overline{\mathit{Receive}}) &  {\displaystyle
      \begin{array}[t]{l}
      \tupleb{\red{\cC\cup\cC'},{\mathsf{rec}({\cC''},\red{\cC'},\red{s},p',\ell,v)\cons h},\tuple{p,s'}}^\tau\\
         \hspace{20ex}  \hoo_{p,\overline{\mathsf{rec}}(\red{\ell})}      
        (\red{\cC'},p',p,\{{\ell},v\}) \comp
           \tuplef{\red{\cC},\cD,h,\tuple{p,\red{s}}}
      \hspace{2ex}\mbox{if \red{$\tau\in\cC''$}} 
     \end{array}
      }\\[4ex]
      
      &  {\displaystyle
      \begin{array}[t]{l}
      \tupleb{\red{\cC\cup\cC'},{\mathsf{rec}({\cC''},\red{\cC'},\red{s},p',\ell,v)\cons h},\tuple{p,s'}}^\tau\\
         \hspace{20ex}  \hoo_{p,\overline{\mathsf{rec}}(\red{\ell})}      
        (\red{\cC'},p',p,\{{\ell},v\}) \comp
           \tupleb{\red{\cC},\cD,h,\tuple{p,\red{s}}}^\tau 
      \hspace{2ex}\mbox{if \red{$\tau\not\in\cC''$}} 
     \end{array}
      }\\[5ex]
      

   (\overline{\mathit{Spawn}}) & {\displaystyle
           \tupleb{\red{\cC},\cD,\mathsf{spawn}(\red{s},\red{p'})\cons h,
           \tuple{p,s'}}^\tau\comp \tupleb{\red{\cC},\emptyset,\nil,\tuple{\red{p'},s_0}}^\tau 
          \hoo_{p,\overline{\mathsf{spawn}}(\red{p'})}
       \tupleb{\red{\cC},\cD,h,\tuple{p,\red{s}}}^\tau 
      }\\[2ex]

& {\displaystyle
   \begin{array}[t]{l}
           \tupleb{{\cC},\cD,\mathsf{spawn}({s},\red{p'})\cons h,
           \tuple{p,s'}}^\tau\comp \tuplef{\cC',\cD',h',\tuple{\red{p'},s''}}\\
          \hspace{20ex}
          \hoo
           \tupleb{{\cC},\cD,\mathsf{spawn}({s},\red{p'})\cons h,
           \tuple{p,s'}}^\tau\comp \tupleb{\cC',\cD',h',\tuple{\red{p'},s''}}^\tau
     \end{array}
      }\\[5ex]
      
    (\overline{\mathit{Check}}) & {\displaystyle
        \tupleb{\cC,\cD,\mathsf{check}(\red{\tau},s)\cons h,\tuple{p,s'}}^\red{\tau}
       \hoo_{p,\overline{\mathsf{check}}(\red{\tau})}
       \tuplef{\cC\setminus\{\tau\},\cD,h,\tuple{p,s}} 
      }
      \\[2ex]  

& {\displaystyle
        \tupleb{\cC,\cD,\mathsf{check}(\red{\tau'},\red{s})\cons h,\tuple{p,s')}}^\red{\tau}
       \hoo_{p,\overline{\mathsf{check}}(\red{\tau'})}
       \tupleb{\cC\setminus\{\tau'\},\cD,h,\tuple{p,\red{s}}}^{\tau}
       \hspace{1.5ex}\mbox{if \red{$\tau\neq\tau'$}} 
      }\\[2ex]

 (\overline{\mathit{Commit}}) & {\displaystyle
   \begin{array}[t]{l}
        \tupleb{\cC,\cD,\mathsf{commit}(\red{\tau'},\red{s})\cons h,\tuple{p,s')}}^\red{\tau}\\
        \hspace{20ex}\hoo_{p,\overline{\mathsf{commit}}(\red{\tau'})}
       \tupleb{\cC\cup\{\red{\tau'}\},\cD,h,\tuple{p,\red{s}}}^{\tau}
       ~\mbox{if}~\tuple{\red{\tau'},\_,\_}\not\in \cD
       \end{array}       
      }\\[4ex]
      
      & {\displaystyle
   \begin{array}[t]{l}
        \tupleb{\cC,\cD\cup\{\tuple{\red{\tau'},h',P}\},\mathsf{commit}(\red{\tau'},\red{s})\cons h,\tuple{p,s')}}^\red{\tau}\\
       \hspace{38ex}\hoo_{p,\overline{\mathsf{commit}}(\red{\tau'})}
       \tupleb{\cC\cup\{\red{\tau'}\},\cD,h,\tuple{p,\red{s}}}^{\tau}
       \end{array}
      }\\
      
  \end{array}
  $\\[3ex]
  
  \begin{minipage}{.9\linewidth}
  \blue{\em (*) We assume the 
  side condition $\tau\in\cC$ holds
  in all rules.}\\
  \blue{\em (**) The second rule of
  $\overline{\mathit{Send}}$ only applies when 
  the message tagged with $\ell$ has been received
  by $p'$ according to history $h'$.}
  \end{minipage}
\caption{Rollback recovery semantics: backward rules} 
\label{fig:rollback-semantics-backward}
\vspace{-3ex}
\end{figure}

Let us now present the backward rules of the rollback semantics,
which are shown in Figure~\ref{fig:rollback-semantics-backward}.

First, 
rule $\overline{\mathit{Seq}}$ applies when the history is headed
by a term of the form $\mathsf{seq}(s)$. 
It simply removes this element from the history and recovers state $s$. 

Rule $\overline{\mathit{Send}}$ distinguishes two cases.
If the message with tag $\ell$ is a floating message (so it has not
been \emph{received}), then we remove the message from the
system and recover the saved state. Otherwise (i.e., 
the message has been consumed by the target process $p'$), 
the rollback mode is propagated to process $p'$, which
will go backwards up to the receiving of the message;
once the floating message is back into the system, 
the first rule applies.

Rule $\overline{\mathit{Receive}}$ also distinguishes two cases.
In both of them, the message is put back into the system as a floating
message and the recorded state is recovered. They differ in that the
first rule considers the case where $\tau$ is a forced checkpoint
introduced by the received message. In this case, we undo the step
and the process resumes its forward computation.\footnote{Here,
we assume that rule $\overline{\mathit{Send}}$ has a higher
priority than rule $\overline{\mathit{Receive}}$, so once a message
is put back into the network, the corresponding message sending
is undone (rather than being received again).} 
Otherwise (i.e., $\tau$ was introduced somewhere else), 
we undo the step but keep the rollback mode for the process.

Rule $\overline{\mathit{Spawn}}$ proceeds in a similar way
as rule $\overline{\mathit{Send}}$: 
if the spawned process is already
in its initial state with an empty history, it is simply removed from the
system. Otherwise, the reversible mode is propagated to the 
spawned process $p'$ (note that if $p'$ is already in reversible
mode, rule $\overline{\mathit{Spawn}}$ does not apply).

Rule $\overline{\mathit{Check}}$ applies when we reach
a checkpoint in the process' history. If the checkpoint has the same
identifier of the initial rollback operator, $\tau$, the job is done and
the process resumes its forward computation 
after undoing one last step.\footnote{We note that,
in its current formulation, we would recover the state immediately
before the checkpoint and, then, would perform the same actions---up
to the nondeterminism of the language---. If the goal was to
implement a safer try\_catch 
(as illustrated in Section~\ref{sec:intro}),
then we could slightly modify the rules so that, when the
rollback is done, we update the recovered state by replacing
the next expression to be evaluated by the expression 
after the call to rollback
(that in $s'$ in rule $\underline{\mathit{Rollback}}$).
We leave this particular extension as future work.}
Otherwise (i.e., the checkpoint in the history has a different identifier, 
$\tau'$), we undo the step, also removing $\tau'$ from the set
of active checkpoints, but keep the rollback mode.

Finally, rule $\overline{\mathit{Commit}}$ considers two
cases: either the commit has been executed (and, thus, the
rollback will eventually undo the associated check too), or
the commit was delayed (see Example~\ref{ex3} below).

\begin{example} \label{ex3}
Consider again the program in Example~\ref{ex2}, where we
now switch the arguments of $\mathsf{commit}$ and 
$\mathsf{rollback}$ in process $p_1$
in order to illustrate the use of delayed commits ($p_2$ and $p_3$ remain the same as before):\footnote{Furthermore,
terms $\mathsf{send}$ and $\mathsf{rec}$ now include message tags
instead of values.}
	\[
	\begin{array}{l@{~~~~~~~~~~~~}l@{~~~~~~~~~~~~}l}
	 \mbox{\bf proc}~p_1 & \mbox{\bf proc}~p_2 & \mbox{\bf proc}~p_3 \\\hline
	 p_2\!\leftarrow\!\mathsf{spawn}()
	 & p_3\!\leftarrow\!\mathsf{spawn}()
	 & \mathsf{rec}(\ell_2) \\
	 
	 \mathsf{send}(p_2,\ell_1)
	 & \mathsf{rec}(\ell_1)
	 & \mathsf{send}(p_2,\ell_4) \\

	 \red{\tau_1\!\leftarrow\!\mathsf{check}}
	 & \mathsf{send}(p_3,\ell_2)
	 & \mathsf{rec}(\ell_6) \\
	 
	 \mathsf{send}(p_2,\ell_3) 
	 & \mathsf{rec}(\ell_3)
	 & \mathsf{send}(p_2,\ell_7) \\

     \red{\tau_2\!\leftarrow\!\mathsf{check}}
	 & \mathsf{rec}(\ell_4) \\

     \mathsf{send}(p_2,\ell_5)
	 & \mathsf{rec}(\ell_5) \\

	 \red{\mathsf{commit}(\tau_1)}
	 & \mathsf{send}(p_3,\ell_6) \\

    \red{\mathsf{rollback}(\tau_2)}
    & \mathsf{rec}(\ell_7) \\
	\end{array}
	\]
In this case, the sequence of configurations of $p_1$ would be as
follows:\footnote{For clarity, we omit some of the arguments 
of history items and some other information which is not 
relevant for the example.}
\[
\begin{array}{l}
  \tuplef{\emptyset,\emptyset,\blue{\nil},\tuple{p_1,s[p_2\!\leftarrow\!\mathsf{spawn}]}} \\

  \hoo  \tuplef{\emptyset,\emptyset,\blue{[\mathsf{spawn}(p_2)]},\tuple{p_1,s[\mathsf{send}(p_2,\ell_1)]}}\\

  \hoo  \tuplef{\emptyset,\emptyset,\blue{[\mathsf{send}(p_2,\ell_1),\mathsf{spawn}(p_2)]},\tuple{p_1,s[\red{\tau_1\!\leftarrow\!\mathsf{check}}]}}\\

  \hoo  \tuplef{\{\tau_1\},\emptyset,\blue{[\mathsf{check}(\tau_1),\mathsf{send}(p_2,\ell_1),\mathsf{spawn}(p_2)]},\tuple{p_1,s[\mathsf{send}(p_2,\ell_3)]}}\\

  \hoo  \tuplef{\{\tau_1\},\emptyset,\blue{[\mathsf{send}(p_2,\ell_3),\mathsf{check}(\tau_1),\mathsf{send}(p_2,\ell_1),
  \ldots]},\tuple{p_1,s[\red{\tau_2\!\leftarrow\!\mathsf{check}}]}}\\

  \hoo  \tuplef{\{\tau_1,\tau_2\},\emptyset,\blue{[\mathsf{check}(\tau_2),\mathsf{send}(p_2,\ell_3),\mathsf{check}(\tau_1),
  \ldots]},\tuple{p_1,s[\mathsf{send}(p_2,\ell_5)]}}\\

  \hoo  \tuplef{\{\tau_1,\tau_2\},\emptyset,\blue{[\mathsf{send}(p_2,\ell_5),\mathsf{check}(\tau_2),\mathsf{send}(p_2,\ell_3),
  \ldots]},
  \tuple{p_1,s[\red{\mathsf{commit}(\tau_1)}]}}\\

  \hoo  \tuplef{\{\tau_2\},\{\tuple{\tau_1,h,\{p_2\}}\},\blue{[\mathsf{commit}(\tau_1),\mathsf{send}(p_2,\ell_5),
  \ldots]},\tuple{p_1,s[\red{\mathsf{rollback}(\tau_2)}]}}\\

  \hoo  \tupleb{\{\tau_2\},\{\tuple{\tau_1,h,\{p_2\}}\},\blue{[\mathsf{commit}(\tau_1),\mathsf{send}(p_2,\ell_5),
  \ldots]},\tuple{p_1,s[\ldots]}}^{\tau_2}\\

  \hoo  \tupleb{\{\tau_1,\tau_2\},\emptyset,\blue{[\mathsf{send}(p_2,\ell_5),\mathsf{check}(\tau_2),\mathsf{send}(p_2,\ell_3),\ldots
  ]},\tuple{p_1,s[\ldots]}}^{\tau_2}\\

  \hoo \tupleb{\{\tau_1,\tau_2\},\emptyset,\blue{[\mathsf{check}(\tau_2),\mathsf{send}(p_2,\ell_3),
  \mathsf{check}(\tau_1),\ldots
  ]},\tuple{p_1,s[\ldots]}}^{\tau_2}\\

  \hoo \tuplef{\{\tau_1\},\emptyset,\blue{[\mathsf{send}(p_2,\ell_3),
  \mathsf{check}(\tau_1),\mathsf{send}(p_2,\ell_1),\mathsf{spawn}(p_2)
  ]},\tuple{p_1,s[\ldots]}}\\
  
  \hoo \ldots

\end{array}
\]
Here, the call $\mathsf{commit}(\tau_1)$ cannot be executed
since the last checkpoint of process $p_1$ is $\tau_2$.
Therefore, it is added as a delayed checkpoint. Then, we have
a call $\mathsf{rollback}(\tau_2)$ which undo the last steps of
$p_1$ (as well as some steps in $p_2$ and $p_3$ in order to
keep causal consistency, which we do not show for simplicity).
\end{example}

\subsubsection{Soundness.}

In the following, we assume a \emph{fair} selection strategy for 
processes, so that each process is eventually reduced. 
Furthermore, we only consider \emph{well-defined}
derivations where the calls $\mathsf{commit}(\tau)$ and 
$\mathsf{rollback}(\tau)$ can only be made by the same
process that created the checkpoint $\tau$, and 
a process can only have one action 
for every checkpoint $\tau$, either $\mathsf{commit}(\tau)$
or $\mathsf{rollback}(\tau)$, but not both.

Soundness is then proved by projecting the configurations of the
rollback semantics to configurations of either the standard
semantics (function \emph{sta}) or a \emph{pure} reversible
semantics (function \emph{rev}).
Then, we prove that every step under the rollback semantics
has a counterpart either under the standard or under the 
reversible semantics, after applying the corresponding 
projections. Formally,\footnote{We denote 
by $\to^=$ the reflexive closure
of a binary relation $\to$, i.e., $(\to^=) = (\to \cup =)$.
We consider the reflexive closure in the claim of 
Theorem~\ref{th:sound2} since some steps under the rollback
semantics have no counterpart under the standard or
reversible semantics. In these cases, the projected configurations
remain the same.}

\begin{theorem} \label{th:sound2}
	Let $d$ be a well-defined derivation under the rollback 
	semantics. Then, for each step $S \hoo S'$ in $d$ we have
	either $sta(S) \boo^= sta(S')$ or  $rev(S) \rlh^= rev(S')$.
\end{theorem}
We have also proved that every computation
between a checkpoint and the corresponding rollback is indeed
reversible for well-defined derivations.
See the
Appendix  
for the technical details.

We leave the study of other interesting 
results of our rollback semantics 
(e.g., minimality and some partial completeness) for future work.

\section{Related Work}\label{sec:relwork}

There is abundant literature on checkpoint-based
rollback recovery to improve fault tolerance (see, e.g., the survey
by Elnozahy \emph{et al} \cite{EAWJ02} and references there in).
In contrast to most of these approaches, our distinctive features are
the extension of the underlying  language with
\emph{explicit} 
operators for rollback recovery, the automatic generation
of forced checkpoints (somehow similarly to communication-induced
checkpointing \cite{TW08}), and the use of a reversible semantics.
Also, we share some
similarities with the checkpointing technique for fault-tolerant
distributed computing of \cite{FV05,KFV14}, although the aim is
in principle different: 
their goal is the definition of a new programming
model where globally consistent checkpoints can be created
(rather than extending an existing message-passing programming 
language with explicit operators for rollback recovery).
Indeed, the use of some form of reversibility is mentioned 
in \cite{FV05} as future work.

The idea of using reversible computation for 
rollback recovery is not new. E.g., 
Perumalla and Park \cite{PP14} already suggested it as
an alternative to other traditional techniques based on
checkpointing. In contrast to our work,
the authors focus on empirically analyzing the trade-off between
fault tolerance based on checkpointing and on 
reversible computation (i.e., memory vs run time), using
a particular example (a particle collision application).
Moreover, since the application is already reversible,
no Landauer embedding is required.

The introduction of a rollback construct in a causal-consistent
concurrent formalism can be traced back to
\cite{GLM14,GLMT15,LMSS11,LLMS12,LMS16}. In these works, 
however, the authors focus on a different formalism and, 
moreover, no explicit checkpointing operator is considered.
These ideas are then transferred to an Erlang-like language
in \cite{NPV16b,LNPV18jlamp}, where an explicit checkpoint
operator is introduced. However, in contrast to our work,
all actions are recorded into a history (i.e., it has no way of
turning the reversible mode off).
In other words, a checkpoint is just a mark in the execution,
but it is not propagated to other processes (as our 
forced checkpoints) and cannot be removed (as a call
to $\mathsf{commit}$ does in our approach). 
More recent formulations of the reversible semantics for 
an Erlang-like language include \cite{LPV19,LPV21,LSZ19,LM20},
but the checkpoint operator has not been considered 
(the focus is on reversible debugging).

The standard 
semantics in Figure~\ref{fig:standard-semantics} is trivially 
equivalent to that considered 
in \cite{LPV21} except for some minor details:
First, 
we follow the simpler and more elegant formulation
of \cite{LSZ19}. For instance, following the
style of \cite{LPV21}, rule \emph{Send}  would have the 
following form:
\[
{\displaystyle
      \frac{s \arro{\mathsf{send}(p',v)}
        s' }{\Gamma;\tuple{p,s} \comp \Pi 
       \boo_{p,\mathsf{send}(\k)}
    \Gamma\cup\{(p,p',v)\};\tuple{p,s'}\comp \Pi}
      }
\]
In this case, messages are stored in a global mailbox, $\Gamma$,
and an expression like ``$\tuple{p,s}\comp\Pi$'' represents all
the processes in the system, 
i.e., $\tuple{p,s}$ is a distinguished process 
(where reduction applies) and $\Pi$ is
the parallel composition of the remaining processes.
In contrast, we have \emph{floating}
messages and select a process to be
reduced by applying (repeteadly) rule \emph{Par}. 
The possible reductions,
though, are the same in both cases.
There are other, minor differences, like considering a rule
to deal with the predefined function \emph{self} (which returns
the pid of a process), and representing a state by a pair $\theta,e$
(environment,expression)

Another difference with the reversible semantics in 
\cite{LNPV18jlamp,LPV21} is that we consider a single
transition relation for systems ($\hoo$). This relation
aims at modeling an actual execution in which a process
proceeds normally forwards but a call to $\mathsf{rollback}$
forces it to go backwards temporarily 
(a situation that can be propagated
to other processes in order to be causally consistent). 
In contrast, \cite{LNPV18jlamp,LPV21} considers 
first an \emph{uncontrolled} semantics ($\rh$ and $\lh$)
which models \emph{all} possible forward and backward 
computations. Then, a \emph{controlled} semantics is
defined on top of it to drive the steps of the 
uncontrolled semantics in order to satisfy both
replay and rollback requests. 

On a different line of work, 
Vassor and Stefani \cite{VS18} formally studied the
relation between rollback recovery and causal-consistent
reversible computation. In particular, they consider the
relation between a distributed checkpoint/rollback scheme based
on (causal) logging (Manetho \cite{EZ92}) 
and a causally-consistent reversible version of $\pi$-calculus 
with a rollback operator  \cite{LMSS11}. 
Their main conclusion is that the latter
can simulate the rollback recovery strategy of Manetho.
Our aim is somehow similar, since we also simulate a 
checkpoint-based rollback recovery strategy using a 
reversible semantics, but there are also some 
significant differences: the considered language is different
(a variant of $\pi$-calculus vs an Erlang-like language),
they only consider a fixed number of processes
(while we accept dynamic process spawning) and,
moreover, no explicit operators are considered
(i.e., our approach is more oriented to introduce a new
programming feature rather than proving a theoretical result).

Very recently, Mezzina, Tiezzi and Yoshida \cite{MTY23} 
introduced a rollback recovery
strategy for session-based programming. 
Besides considering
a different setting (a variant of $\pi$-calculus), their approach
is also limited to a fixed number of parties (no dynamic processes
can be added at run time), and nested checkpoints are not
allowed. Furthermore, the checkpoints of \cite{MTY23} are
not automatically propagated to other causally consistent
processes (as our forced checkpoints); rather, they introduce a
\emph{compliance check} at the type level to prevent 
undesired situations.

Our work also shares some similarities with \cite{SKM17}, which
presents a hybrid model combining message-passing concurrency 
and software transactional memory. 
However, the underlying 
language is different and, moreover, their
transactions cannot include process spawning 
(which must be delayed).

Finally, Fabbretti, Lanese and Stefani \cite{FLS23} introduced a
calculus to formally model distributed systems subject to crash 
failures, where recovery mechanisms can be encoded by
a small set of primitives. This work can be seen
as a reworking and extension of the previous work by
Francalanza and Hennessy \cite{FH08}. 
Here, a variant of $\pi$-calculus is considered. Furthermore,
the authors focus on crash recovery without relying on
a form of checkpointing, in contrast to our approach.

\section{Conclusions and Future Work}\label{sec:conc} 

In this work, we have defined a rollback-recovery strategy
for a message-passing concurrent programming language 
without the need for a central coordination.
For this purpose, we have extended the underlying language 
with three explicit operators: 
$\mathsf{check}$, $\mathsf{commit}$, and $\mathsf{rollback}$. 
Our approach
is based on a reversible semantics where
every process may go both forwards and backwards (during
a rollback). Checkpoints are automatically propagated 
to other processes so that backward computations are
causally consistent. 
The ability to turn the reversible mode on/off is useful not
only to model rollback recovery, but can also constitute
the basis of a safer try\_catch 
(as illustrated in Section~\ref{sec:intro}) and
a \emph{selective} reversible debugging scheme, 
where only some computations---those of interest---are traced,
thus making it easier to scale to larger applications.

As for future work, we will consider the definition of a 
\emph{shorcut} version of the rollback semantics where only
the state in a checkpoint is recorded (rather than all the states
between a checkpoint and the corresponding commit/rollback)
so that a rollback recovers the saved state in one go.
This extension 
will be essential to make our approach feasible in practice.
In the context of Erlang, a prototype implementation of the 
proposed operators ($\mathsf{check}$, $\mathsf{commit}$, and $\mathsf{rollback}$) could be carried out through a 
program instrumentation. 
It will likely require introducing a \emph{wrapper} for each process
in order to record the process' history, turning the 
reversible mode on/off, propagating forced checkpoints
and commits, etc. For this purpose, one could explore the
use of the run-time monitors of \cite{FMT18}, which play a similar
role in their scheme for reversible choreographies.

\subsubsection*{Acknowledgements.}

The author would like to thank Ivan Lanese 
and Adri\'an Palacios for their useful remarks 
and discussions on a preliminary version of
this work. 
I would also like to thank the anonymous reviewers 
and the participants 
of FACS 2023 for their suggestions to improve this paper.

\bibliographystyle{splncs04}

\begin{thebibliography}{10}
\providecommand{\url}[1]{\texttt{#1}}
\providecommand{\urlprefix}{URL }
\providecommand{\doi}[1]{https://doi.org/#1}

\bibitem{ACGKKKLMMNPPPUV20}
Aman, B., et~al.: Foundations of reversible computation. In: Ulidowski, I.,
  Lanese, I., Schultz, U.P., Ferreira, C. (eds.) Reversible Computation:
  Extending Horizons of Computing - Selected Results of the {COST} Action
  {IC1405}, Lecture Notes in Computer Science, vol. 12070, pp. 1--40. Springer
  (2020). \doi{10.1007/978-3-030-47361-7\_1}

\bibitem{DK04}
Danos, V., Krivine, J.: Reversible communicating systems. In: CONCUR. LNCS,
  vol.~3170, pp. 292--307. Springer (2004)

\bibitem{EAWJ02}
Elnozahy, E.N., Alvisi, L., Wang, Y., Johnson, D.B.: A survey of
  rollback-recovery protocols in message-passing systems. {ACM} Comput. Surv.
  \textbf{34}(3),  375--408 (2002)

\bibitem{EZ92}
Elnozahy, E.N., Zwaenepoel, W.: {Manetho: Transparent Rollback-Recovery with
  Low Overhead, Limited Rollback, and Fast Output Commit}. {IEEE} Trans.
  Computers  \textbf{41}(5),  526--531 (1992). \doi{10.1109/12.142678}

\bibitem{erlang}
{Erlang} website. URL: \url{https://www.erlang.org/} (2021)

\bibitem{FLS23}
Fabbretti, G., Lanese, I., Stefani, J.B.: A behavioral theory for crash
  failures and erlang-style recoveries in distributed systems. Tech. Rep.
  RR-9511, INRIA (2023), \url{https://hal.science/hal-04123758}

\bibitem{ErlangFAQ}
Frequently {A}sked {Q}uestions about {E}rlang. Available at
  \url{http://erlang.org/faq/academic.html} (2018)

\bibitem{FV05}
Field, J., Varela, C.A.: Transactors: a programming model for maintaining
  globally consistent distributed state in unreliable environments. In:
  Palsberg, J., Abadi, M. (eds.) Proceedings of the 32nd {ACM} {SIGPLAN-SIGACT}
  Symposium on Principles of Programming Languages ({POPL} 2005). pp. 195--208.
  {ACM} (2005)

\bibitem{FH08}
Francalanza, A., Hennessy, M.: A theory of system behaviour in the presence of
  node and link failure. Inf. Comput.  \textbf{206}(6),  711--759 (2008).
  \doi{10.1016/j.ic.2007.12.002}

\bibitem{FMT18}
Francalanza, A., Mezzina, C.A., Tuosto, E.: Reversible choreographies via
  monitoring in erlang. In: Bonomi, S., Rivi{\`{e}}re, E. (eds.) Proceedings of
  the 18th {IFIP} {WG} 6.1 International Conference on Distributed Applications
  and Interoperable Systems ({DAIS} 2018), held as part of DisCoTec 2018.
  Lecture Notes in Computer Science, vol. 10853, pp. 75--92. Springer (2018).
  \doi{10.1007/978-3-319-93767-0\_6}

\bibitem{GLM14}
Giachino, E., Lanese, I., Mezzina, C.A.: Causal-consistent reversible
  debugging. In: Gnesi, S., Rensink, A. (eds.) Proceedings of the 17th
  International Conference on Fundamental Approaches to Software Engineering
  ({FASE} 2014). Lecture Notes in Computer Science, vol.~8411, pp. 370--384.
  Springer (2014)

\bibitem{GLMT15}
Giachino, E., Lanese, I., Mezzina, C.A., Tiezzi, F.: Causal-consistent
  reversibility in a tuple-based language. In: Daneshtalab, M., Aldinucci, M.,
  Lepp{\"{a}}nen, V., Lilius, J., Brorsson, M. (eds.) Proceedings of the 23rd
  Euromicro International Conference on Parallel, Distributed, and
  Network-Based Processing, {PDP} 2015. pp. 467--475. {IEEE} Computer Society
  (2015)

\bibitem{GLMMPUV23}
Gl{\"u}ck, R., Lanese, I., Mezzina, C.A., Miszczak, J.A., Phillips, I.,
  Ulidowski, I., Vidal, G.: Towards a taxonomy for reversible computation
  approaches. In: Kutrib, M., Meyer, U. (eds.) Reversible Computation. pp.
  24--39. Springer Nature Switzerland, Cham (2023)

\bibitem{GV21tr}
Gonz\'alez-Abril, J.J., Vidal, G.: {Causal-Consistent Reversible Debugging:
  Improving CauDEr}. Tech. rep., DSIC, Universitat Polit\`ecnica de Val\`encia
  (2020), \url{https://gvidal.webs.upv.es/confs/padl21/tr.pdf}

\bibitem{GV21}
Gonz{\'{a}}lez{-}Abril, J.J., Vidal, G.: {Causal-Consistent Reversible
  Debugging: Improving CauDEr}. In: Morales, J.F., Orchard, D.A. (eds.)
  Proceedings of the 23rd International Symposium on Practical Aspects of
  Declarative Languages ({PADL} 2021). Lecture Notes in Computer Science, vol.
  12548, pp. 145--160. Springer (2021). \doi{10.1007/978-3-030-67438-0\_9}

\bibitem{HBS73}
Hewitt, C., Bishop, P.B., Steiger, R.: A universal modular {ACTOR} formalism
  for artificial intelligence. In: Nilsson, N.J. (ed.) Proceedings of the 3rd
  International Joint Conference on Artificial Intelligence. pp. 235--245.
  William Kaufmann (1973),
  \url{http://ijcai.org/Proceedings/73/Papers/027B.pdf}

\bibitem{KFV14}
Kuang, P., Field, J., Varela, C.A.: Fault tolerant distributed computing using
  asynchronous local checkpointing. In: Boix, E.G., Haller, P., Ricci, A.,
  Varela, C. (eds.) Proceedings of the 4th International Workshop on
  Programming based on Actors Agents {\&} Decentralized Control (AGERE! 2014).
  pp. 81--93. {ACM} (2014)

\bibitem{Lam78}
Lamport, L.: Time, clocks, and the ordering of events in a distributed system.
  Commun.\ {ACM}  \textbf{21}(7),  558--565 (1978). \doi{10.1145/359545.359563}

\bibitem{Lan61}
Landauer, R.: Irreversibility and heat generation in the computing process. IBM
  Journal of Research and Development  \textbf{5},  183--191 (1961)

\bibitem{LM20}
Lanese, I., Medic, D.: A general approach to derive uncontrolled reversible
  semantics. In: Konnov, I., Kov{\'{a}}cs, L. (eds.) 31st International
  Conference on Concurrency Theory, {CONCUR} 2020. LIPIcs, vol.~171, pp.
  33:1--33:24. Schloss Dagstuhl - Leibniz-Zentrum f{\"{u}}r Informatik (2020).
  \doi{10.4230/LIPIcs.CONCUR.2020.33}

\bibitem{LMSS11}
Lanese, I., Mezzina, C.A., Schmitt, A., Stefani, J.: Controlling reversibility
  in higher-order pi. In: Katoen, J., K{\"{o}}nig, B. (eds.) Proceedings of the
  22nd International Conference on Concurrency Theory ({CONCUR} 2011). Lecture
  Notes in Computer Science, vol.~6901, pp. 297--311. Springer (2011)

\bibitem{LMS16}
Lanese, I., Mezzina, C.A., Stefani, J.: Reversibility in the higher-order
  {\(\pi\)}-calculus. Theor. Comput. Sci.  \textbf{625},  25--84 (2016)

\bibitem{LNPV18jlamp}
Lanese, I., Nishida, N., Palacios, A., Vidal, G.: A theory of reversibility for
  {E}rlang. Journal of Logical and Algebraic Methods in Programming
  \textbf{100},  71--97 (2018). \doi{10.1016/j.jlamp.2018.06.004}

\bibitem{LPV19}
Lanese, I., Palacios, A., Vidal, G.: Causal-consistent replay debugging for
  message passing programs. In: P{\'{e}}rez, J.A., Yoshida, N. (eds.)
  Proceedings of the 39th {IFIP} {WG} 6.1 International Conference on Formal
  Techniques for Distributed Objects, Components, and Systems ({FORTE} 2019).
  Lecture Notes in Computer Science, vol. 11535, pp. 167--184. Springer (2019).
  \doi{10.1007/978-3-030-21759-4\_10}

\bibitem{LPV21}
Lanese, I., Palacios, A., Vidal, G.: Causal-consistent replay reversible
  semantics for message passing concurrent programs. Fundam. Informaticae
  \textbf{178}(3),  229--266 (2021). \doi{10.3233/FI-2021-2005}

\bibitem{LSZ19}
Lanese, I., Sangiorgi, D., Zavattaro, G.: Playing with bisimulation in
  {E}rlang. In: Boreale, M., Corradini, F., Loreti, M., Pugliese, R. (eds.)
  Models, Languages, and Tools for Concurrent and Distributed Programming --
  Essays Dedicated to Rocco De Nicola on the Occasion of His 65th Birthday.
  Lecture Notes in Computer Science, vol. 11665, pp. 71--91. Springer (2019).
  \doi{10.1007/978-3-030-21485-2\_6}

\bibitem{LLMS12}
Lienhardt, M., Lanese, I., Mezzina, C.A., Stefani, J.B.: A reversible abstract
  machine and its space overhead. In: Giese, H., Rosu, G. (eds.) Proceedings of
  the Joint 14th {IFIP} {WG} International Conference on Formal Techniques for
  Distributed Systems ({FMOODS} 2012) and the 32nd {IFIP} {WG} 6.1
  International Conference ({FORTE} 2012). Lecture Notes in Computer Science,
  vol.~7273, pp. 1--17. Springer (2012)

\bibitem{MHNHT07}
Matsuda, K., Hu, Z., Nakano, K., Hamana, M., Takeichi, M.: Bidirectionalization
  transformation based on automatic derivation of view complement functions.
  In: Hinze, R., Ramsey, N. (eds.) Proc.\ of the 12th {ACM} {SIGPLAN}
  International Conference on Functional Programming, {ICFP} 2007. pp. 47--58.
  {ACM} (2007)

\bibitem{MTY23}
Mezzina, C.A., Tiezzi, F., Yoshida, N.: Rollback recovery in session-based
  programming. In: Jongmans, S., Lopes, A. (eds.) Proceedings of the 25th
  {IFIP} {WG} 6.1 International Conference on Coordination Models and
  Languages, {COORDINATION} 2023. Lecture Notes in Computer Science, vol.
  13908, pp. 195--213. Springer (2023). \doi{10.1007/978-3-031-35361-1\_11}

\bibitem{Mil80}
Milner, R.: A {C}alculus of {C}ommunicating {S}ystems. Springer LNCS 92 (1980)

\bibitem{NPV16b}
Nishida, N., Palacios, A., Vidal, G.: A reversible semantics for {E}rlang. In:
  Hermenegildo, M., L\'opez-Garc\'{\i}a, P. (eds.) Proceedings of the 26th
  International Symposium on Logic-Based Program Synthesis and Transformation
  (LOPSTR 2016). Lecture Notes in Computer Science, vol. 10184, pp. 259--274.
  Springer (2017). \doi{10.1007/978-3-319-63139-4\_15}

\bibitem{NPV18}
Nishida, N., Palacios, A., Vidal, G.: Reversible computation in term rewriting.
  J. Log. Algebraic Methods Program.  \textbf{94},  128--149 (2018).
  \doi{10.1016/j.jlamp.2017.10.003}

\bibitem{PP14}
Perumalla, K.S., Park, A.J.: Reverse computation for rollback-based fault
  tolerance in large parallel systems - evaluating the potential gains and
  systems effects. Clust. Comput.  \textbf{17}(2),  303--313 (2014).
  \doi{10.1007/s10586-013-0277-4}

\bibitem{SKM17}
Swalens, J., Koster, J.D., Meuter, W.D.: Transactional actors: communication in
  transactions. In: Jannesari, A., de~Oliveira~Castro, P., Sato, Y., Mattson,
  T. (eds.) Proceedings of the 4th {ACM} {SIGPLAN} International Workshop on
  Software Engineering for Parallel Systems, SEPS\@SPLASH 2017. pp. 31--41.
  {ACM} (2017). \doi{10.1145/3141865.3141866}

\bibitem{TW08}
Tsai, J., Wang, Y.: Communication-induced checkpointing protocols and
  rollback-dependency trackability: {A} survey. In: Wah, B.W. (ed.) Wiley
  Encyclopedia of Computer Science and Engineering. John Wiley {\&} Sons, Inc.
  (2008). \doi{10.1002/9780470050118.ecse059}

\bibitem{VS18}
Vassor, M., Stefani, J.B.: {Checkpoint/Rollback vs Causally-Consistent
  Reversibility}. In: Kari, J., Ulidowski, I. (eds.) Reversible Computation.
  pp. 286--303. Springer International Publishing, Cham (2018).
  \doi{978-3-319-99498-7\_20}

\end{thebibliography}

\clearpage
\appendix


\section{Reversible Semantics}
\label{appendix:semantics}

In this section, we present the reversible semantics of
\cite{LPV19,LPV21} but omitting the \emph{replay} component using
an execution trace, which is not relevant in our context.
Furthermore, there are a few, minor differences:
\begin{itemize}
\item First, we consider floating messages and rule \emph{Par} to
lift reductions to a larger system (as in 
\cite{LSZ19,LM20}). 
The resulting semantics is
straightforwardly equivalent but the formulation of the 
rules is simpler.

\item We consider a generic state, $s$, rather than a pair
$\theta,e$ (environment, expression) as in  \cite{LPV19,LPV21}.
This is a simple generalization to improve readability 
but does not affect the behavior of the system rules.

\item Finally, we do not consider a rule for the predefined function
\emph{self()} (that returns the pid of the current process) since it
is not relevant in the context this work (but could be added easily).
\end{itemize}

\begin{figure}[p]
  \[
  \hspace{-3ex}
  \begin{array}{r@{~~}c}
      (\underline{\mathit{Seq}}) & {\displaystyle
      \frac{s \arro{\mathsf{seq}} s'
      }{\tuplef{h,\tuple{p,s}} 
        \rh_{p,\underline{\mathsf{seq}}} 
        \tuplef{{\mathsf{seq}(s)\cons h},\tuple{p,s'}}}
      }\\[3ex]

    (\underline{\mathit{Send}}) & {\displaystyle
      \frac{s \arro{\mathsf{send}(p',v)} s' ~\mbox{and}~ 
      ~{\ell~\mbox{is a fresh symbol}}
      }{\tuplef{h,\tuple{p,s}} 
        \rh_{p,\underline{\mathsf{send}}({\ell})}
         (p,p',\{{\ell},v\}) \comp 
         \tuplef{{\mathsf{send}(s,p',\ell)\cons h},\tuple{p,s'}}}
      }\\[3ex]
      
      (\underline{\mathit{Receive}}) &  {\displaystyle
        \frac{s \arro{\mathsf{rec}(\kappa,cs)}
          s' ~\mbox{and}~ \mathsf{matchrec}(cs,v) = cs_i
         }
          {(p',p,\{{\ell},v\}) \comp
           \tuplef{h,\tuple{p,s}} 
           \rh_{p,\underline{\mathsf{rec}}({\ell})}
           \tuplef{{\mathsf{rec}(s,p',\ell,v)\cons h},\tuple{p,s'[\kappa\leftarrow cs_i]}}}
      }\\[3ex]  
      
   (\underline{\mathit{Spawn}}) & {\displaystyle
        \frac{s \arro{\mathsf{spawn}(\kappa,s_0)} 
          s'~\mbox{and}~ p'~\mbox{is a fresh pid}
          }
          {\tuplef{h,\tuple{p,s}} \rh_{p,\underline{\mathsf{spawn}}(p')}    
           \tuplef{{\mathsf{spawn}(s,p')\cons h},\tuple{p,s'[\kappa\leftarrow p']}}\comp \tuplef{\nil,\tuple{p',s_0}}}
      }\\[3ex]

      (\underline{\mathit{Par}}) & {\displaystyle
      \frac{S_1 \rh_{l} S'_1~~\mbox{and}~~\ell(S'_1)\cap \ell(S_2)=\emptyset}{S_1  \comp S_2 
         \rh_{{l}} S'_1 \comp S_2 
        }
      }      

  \end{array}
  \]
\caption{Reversible semantics: forward rules} \label{fig:reversible-semantics-forward}
\end{figure}

\begin{figure}[p]
  \[
  \hspace{-3ex}
  \begin{array}{r@{~~}c}
    (\overline{\mathit{Seq}}) & {\displaystyle
      \tuplef{\mathsf{seq}({s})\cons h,\tuple{p,s'}}
      \lh_{p,\overline{\mathsf{seq}}} 
      \tuplef{h,\tuple{p,{s}}}
      }
      \\[2ex]
  
    (\overline{\mathit{Send}}) & {\displaystyle
      (p,p',\{{\ell},v\}) \comp 
         \tuplef{\mathsf{send}({s},p',{\ell})\cons h,
         \tuple{p,s'}} \lh_{p,\overline{\mathsf{send}}({\ell})}         
         \tuplef{h,\tuple{p,{s}}}
      }\\[2ex]
        
      (\overline{\mathit{Receive}}) &  {\displaystyle
      \tuplef{{\mathsf{rec}({s},p',\ell,v)\cons h},\tuple{p,s'}}
      \lh_{p,\overline{\mathsf{rec}}({\ell})}      
        (p',p,\{{\ell},v\}) \comp
           \tuplef{h,\tuple{p,{s}}}
      }\\[2ex]
      
   (\overline{\mathit{Spawn}}) & {\displaystyle
           \tuplef{\mathsf{spawn}({s},{p'})\cons h,
           \tuple{p,s'}}\comp \tuplef{\nil,\tuple{{p'},s_0}} 
           \lh_{p,\overline{\mathsf{spawn}}(p')}
       \tuplef{h,\tuple{p,{s}}}
      }\\[1ex]
      
          (\overline{\mathit{Par}}) & {\displaystyle
      \frac{S_1 \lh_{l} S'_1}{S_1  \comp S_2 
         \lh_{{l}} S'_1 \comp S_2 
        }
      }

  \end{array}
  \]
\caption{Reversible semantics: backward rules} \label{fig:reversible-semantics-backward}
\end{figure}

\noindent
The \emph{uncontrolled} reversible semantics is defined in
Figures~\ref{fig:reversible-semantics-forward} and 
\ref{fig:reversible-semantics-backward}. 
The forward rules (Fig.~\ref{fig:reversible-semantics-forward})
do not need any additional explanation since they are perfectly
similar to the rollback recovery rules in 
Figure~\ref{fig:rollback-semantics-forward}.
The main difference comes from the fact that, in this case,
the reversible mode is always on.
As for the backward rules 
(Fig.~\ref{fig:reversible-semantics-backward}), the main difference
w.r.t.\ the rollback recovery rules in 
Figure~\ref{fig:rollback-semantics-backward}
is the fact that they are \emph{uncontrolled}: there is no
rollback operator that drives the backward computation.
Therefore, we only have one rule for each case that can
be only applied when the conditions are met, i.e.,
$\mathsf{spawn}$ can only be undone when the spawned
process is in its initial state and $\mathsf{send}$ can only
be undone when the message is in the system.

The uncontrolled semantics, $\rlh$, can then be defined as the union
of the two transition relations defined in
Figures~\ref{fig:reversible-semantics-forward} and 
\ref{fig:reversible-semantics-backward}, i.e.,
$(\rlh) \: = \: (\rh \cup \lh)$.
The causal consistency of $\rlh$ can be proved analogously
to \cite[Theorem 4.17]{LPV21}.

On top of the uncontrolled semantics, \cite{LNPV18jlamp,LPV21}
introduces a \emph{controlled} semantics that can be used
to \emph{drive} backward computations in order to satisfy
different requests, e.g.,
\begin{itemize}
\item go backwards one (or more) steps;
\item go backwards up to the sending of a given message;
\item go backwards up to the spawning of a given process;
\item etc.
\end{itemize}
Notably, all these requests are carried over in a causal
consistent way, thus they often require other processes to
go backwards too. In particular, in order to perform a 
backward step in a given process $p$, the controlled semantics
proceeds essentially as follows:
\begin{enumerate}
\item If $p$ can perform a backward step under the
uncontrolled semantics, we are done.
\item Otherwise, we distinguish two cases:
\begin{itemize}
\item If the first action in the history of $p$ is a message
sending to process $p'$ and the message is not in the system
(i.e., it has been received), then the controlled semantics 
adds a new request for $p'$ to go backwards up to the
receiving of this message. After $p'$ completes this
request, case (1) above will be applicable (since the message
will be in the system).
\item If the first action in the history of $p$ is a spawn of
process $p'$ and this process is not in its initial state,
then the controlled semantics adds a new request for $p'$
to go backwards up to its initial state. Once $p'$ completes 
the request, case (1) above can be applied.
\end{itemize}
\end{enumerate}
Observe that this is similar to the propagation of the
rollback mode in our semantics (check the second rules of 
$\overline{\mathit{Send}}$ and $\overline{\mathit{Spawn}}$
in Figure~\ref{fig:rollback-semantics-backward}).

\section{Soundness of the Rollback Recovery Semantics}
\label{appendix:soundness}

In the section, we prove the soundness of the 
rollback recovery semantics of 
Figures~\ref{fig:rollback-semantics-forward}
and \ref{fig:rollback-semantics-backward}.
First, we introduce the following auxiliary functions that define
projections from rollback configurations to either standard
or reversible configurations. Function $sta$ takes
a system of the rollback semantics and returns a corresponding 
system of the standard semantics:
\[
sta(S) = \left\{ \begin{array}{ll}
	sta(S_1)\comp sta(S_2) & \mbox{if}~ S = S_1\comp S_2 \\
	\tuple{p,\overline{s}} & \mbox{if}~S=\tuplef{\cC,\cD,h,\tuple{p,s}} \\
	\tuple{p,\overline{s}} & \mbox{if}~S=\tupleb{\cC,\cD,h,\tuple{p,s}}^\tau \\
	(p,p',v) & \mbox{if}~S=(\cC,p,p',\{\ell,v\}) \\
\end{array}
\right.
\]
where $\overline{s}$ replaces from $s$ all occurrences of
the rollback operators ($\mathsf{check}$, $\mathsf{commit}$,
and $\mathsf{rollback}$) and checkpoint identifiers (if any)
by an arbitrary constant ``ok'' (an atom, using Erlang 
terminology).
Function $rev$ does a similar projection to a system of the
reversible semantics:
\[
rev(S) = \left\{ \begin{array}{ll}
	rev(S_1)\comp rev(S_2) & \mbox{if}~ S = S_1\comp S_2 \\
	\tuplef{{r}(h),\tuple{p,\overline{s}}} & \mbox{if}~S=\tuplef{\cC,\cD,h,\tuple{p,s}} \\
	\tuplef{{r}(h),\tuple{p,\overline{s}}} & \mbox{if}~S=\tupleb{\cC,\cD,h,\tuple{p,s}}^\tau \\
	(p,p',\{\ell,v\}) & \mbox{if}~S=(\cC,p,p',\{\ell,v\}) \\
\end{array}
\right.
\]
where function $r$ is used to clean up some 
history elements as follows:
\[
r(h) = \left\{ \begin{array}{ll}
	\mathsf{seq}(s)\cons r(h') 
	& \mbox{if}~ h = \mathsf{seq}(s)\cons h' \\
	\mathsf{send}(s,p',\ell)\cons r(h') 
	& \mbox{if}~ h = \mathsf{send}(s,p',\ell)\cons h' \\
	\mathsf{rec}(s,p',\ell,v)\cons r(h') 
	& \mbox{if}~ h = \mathsf{rec}(\cC,\cC',s,p',\ell,v)\cons h' \\
	\mathsf{spawn}(s,p')\cons r(h') 
	& \mbox{if}~ h = \mathsf{spawn}(s,p')\cons h' \\
	 r(h') 
	& \mbox{if}~ h = \mathsf{check}(\tau,s)\cons h' \\
	r(h') & \mbox{if}~ h = \mathsf{commit}(\tau,s)\cons h' \\
\end{array}
\right.
\]
Functions $\mathit{sta}$ and $\mathit{rev}$ 
are extended to derivations in the obvious way.

\begin{definition}[reachable system]
	Let $S$ be a system. We say that $S$ is a reachable
	system 
	if there exists a derivation of the form
	$\tuplef{\emptyset,\emptyset,\nil,\tuple{p,s}} \hoo^\ast S$ 
	under the rollback semantics, for some pid $p$ and state $s$.
\end{definition}
In the following, we consider some minimal requirements
on derivations under the rollback semantics 
in order to be well-defined:
\begin{itemize}
\item The calls $\mathsf{commit}(\tau)$ and 
$\mathsf{rollback}(\tau)$ can only be made by the same
process that created the checkpoint $\tau$ (i.e., that
called $\mathsf{check}$ and was reduced to $\tau$).

\item Every call to either $\mathsf{commit}(\tau)$ or
$\mathsf{rollback}(\tau)$ must be preceded by a call
to $\mathsf{check}$ returning $\tau$.

\item A process can only have one action 
for every checkpoint $\tau$, either $\mathsf{commit}(\tau)$
or $\mathsf{rollback}(\tau)$, but not both.
This condition can easily be verified from the history
of a process. Note that, if a rollback (associated to a different 
checkpoint) undoes a commit,
it is removed from the history of the process and, thus, 
a new commit/rollback would be possible. 
For instance, at the end of the derivation shown in
Example~\ref{ex3}, a new call $\mathsf{commit}(\tau_1)$
or $\mathsf{rollback}(\tau_1)$ would be possible, since
the initial call $\mathsf{commit}(\tau_1)$ has been undone
by the call $\mathsf{rollback}(\tau_2)$.
\end{itemize}
More formally,

\begin{definition}[well-defined derivation] \label{def:welldefined}
	A derivation $d = (S_0 \hoo S_1 \hoo \ldots \hoo S_n)$ 
	under the rollback semantics 
	is well-defined if the following conditions hold:
	\begin{enumerate}
	\item $S_0$ is a reachable configuration;
	\item every reduction step $S_i \hoo_{p,\underline{\mathsf{commit}}(\tau)} S_{i+1}$ 
	in $d$ is preceded by a reduction step $S_j \hoo_{p,\overline{\mathsf{check}}(\tau)} S_{j+1}$, $j<i$;
	\item every reduction step $S_i \hoo_{p,\underline{\mathsf{rollback}}(\tau)} S_{i+1}$ 
	in $d$ is preceded by a reduction step $S_j \hoo_{p,\overline{\mathsf{check}}(\tau)} S_{j+1}$, $j<i$;
	\item no process history in $d$ may have both
	$\mathsf{commit}(\tau)$ and $\mathsf{rollback}(\tau)$
	at the same time.
	\end{enumerate}
\end{definition}
For instance, it is easy to prove that derivations are well-defined
when the new operators are used to improve try\_catch as 
proposed in the introduction\\

$
\mathtt{try}~\red{T = \mathsf{check}},~\red{X=\;} e,~\red{\mathsf{commit}(T)},~\red{X}~\mathtt{catch}~\_:\_ \rightarrow \red{\mathsf{rollback}(T)},~e'~\mathtt{end}
$\\

\noindent
First, we prove that our rollback semantics is indeed a
conservative extension of the standard semantics:

\begin{theorem} \label{th:sound1}
	Let $d$ be a well-defined derivation under the rollback 
	semantics where only forward rules are applied. 
	Then, $sta(d)$ is a derivation under the
	standard semantics.
\end{theorem}

\begin{proof}
	Rule \underline{\emph{Par}} is the same in both semantics. 
	Now, we consider an arbitrary (forward) 
	step of $d$, $S \hoo_{p,a} S'$
	and prove that either there exists a corresponding step 
	$sta(S) \boo_{p,a} sta(S')$ in $sta(d)$ or
	$sta(S) = sta(S')$.
	
	If the considered step applies one of the first four rules 
	(\underline{\emph{Seq}},
	\underline{\emph{Send}}, \underline{\emph{Receive}},
	and \underline{\emph{Spawn}}), the claim follows trivially
	since the component $\tuple{p,s}$ of the rollback configurations
	is the same in both semantics.
	
	Let us now consider rule \underline{\emph{Check}} and a 
	step of the form\\[-1ex]
	
	$\tuplef{\cC,\cD,h,\tuple{p,s[\mathsf{check}]}}
	\hoo_{p,\underline{\mathsf{check}}(\tau)} 
	\tuplef{\cC\cup\{\tau\},\cD,\mathsf{check}(\tau,s)\cons h,\tuple{p,s[\tau]}}
	$\\[-1ex]	
	
	\noindent
	Here, we have that $sta(\tuplef{\cC,\cD,h,\tuple{p,s[\mathsf{check}]}})=
	sta(\tuplef{\cC\cup\{\tau\},\cD,\mathsf{check}(\tau,s)\cons h,\tuple{p,s[\tau]}})
	= \tuple{p,\overline{s}[ok]}$ since both the call to
	$\mathsf{check}$ and the checkpoint identifier are
	replaced with $ok$, thus the claim follows.
	
	If the applied rule is one of the \underline{\emph{Commit}} rules,  
	then the step has either the form \\[-1ex]
	
	$\tuplef{\cC,\cD,h,\tuple{p,s[\mathsf{commit}(\tau)]}}
	\hoo_{p,\underline{\mathsf{commit}}(\tau)} 
	\tuplef{\cC\setminus\{\tau\},\cD,\mathsf{commit}(\tau,s)\cons h,\tuple{p,s[ok]}}
	$\\[-1ex]	
	
	\noindent
	or\\[-1ex]
	
   $\begin{array}{l}
   \tuplef{\cC,\cD,h,\tuple{p,s[\mathsf{commit}(\tau)]}}\\
	\hspace{14ex}\hoo_{p,\underline{\mathsf{commit}}(\tau)} 
	\tuplef{\cC\setminus\{\tau\},\cD\cup\{\tuple{\tau,h,P}\},\mathsf{commit}(\tau,s)\cons h,\tuple{p,s[ok]}}
   \end{array}
	$\\[-1ex]	

	\noindent
	Trivially, the application of function $sta$ returns the same
	configuration, $\tuple{p,s[ok]}$ for all four configurations
	above, and the claim holds.
		
	The case of rule \underline{\emph{Delay}} is trivial too
	since the standard component of the configurations is
	identical.

	Finally, if the applied rule is \underline{\emph{Rollback}},  
	then the step has the form \\[-1ex]
	
	$\tuplef{\cC,\cD,h,\tuple{p,s[\mathsf{rollback}(\tau)]}}
	\hoo_{p,\underline{\mathsf{rollback}}(\tau)} 
	\tupleb{\cC,\cD,h,\tuple{p,s[ok]}}^\tau
	$\\[-1ex]
	
	\noindent
	and the claim follows trivially since\\[-1ex] 
	
	$sta(\tuplef{\cC,\cD,h,\tuple{p,s[\mathsf{rollback}(\tau)]}})
	= sta(\tupleb{\cC,\cD,h,\tuple{p,s[ok]}})
	= \tuple{p,\overline{s}[ok]}$. 
	\qed
\end{proof}
Now, we state the soundness of our rollback semantics;
namely, we prove that each reduction step under the
rollback semantics (Figures~\ref{fig:rollback-semantics-forward} and
\ref{fig:rollback-semantics-backward}) 
has a corresponding reduction step
either under the standard semantics 
(Figure~\ref{fig:standard-semantics}) or under the
reversible semantics (Figures~\ref{fig:reversible-semantics-forward}
and \ref{fig:reversible-semantics-backward}).

In the following, we denote by $\to^=$ the reflexive closure
of a binary relation $\to$, i.e., $(\to^=) = (\to \cup =)$.

\mbox{}\\
\textbf{Theorem~\ref{th:sound2}.}
\emph{Let $d$ be a well-defined derivation under the rollback 
	semantics. Then, for each step $S \hoo S'$ in $d$ we have
	either $sta(S) \boo^= sta(S')$ or  $rev(S) \rlh^= rev(S')$.}

\begin{proof}
	First, let us consider that the selected process in $S$ 
	has an empty set of active checkpoints and, moreover,
	one of the first four rules (\underline{\emph{Seq}},
	\underline{\emph{Send}}, \underline{\emph{Receive}},
	\underline{\emph{Spawn}}) is applied.
	 Then, the claim follows by Theorem~\ref{th:sound1}.

	Let us now consider the remaining cases. 
	If the applied rule is \underline{\emph{Seq}}, 
	\underline{\emph{Send}}, \underline{\emph{Receive}}, or
	\underline{\emph{Spawn}}, and the set of active
	checkpoints is not empty, the claim follows trivially by
	definition of function $rev$ and the fact that the 
	transition rules are identical if we ignore the sets of
	active checkpoints. 
	Consider now the following cases:
	\begin{itemize}
	\item Assume that the applied rule is \underline{\emph{Check}}.
	Then, the reduction step has the form
	$\tuplef{\cC,\cD,h,\tuple{p,s[\mathsf{check}]}}
	\hoo_{p,\underline{\mathsf{check}}(\tau)}
	\tuplef{\cC\cup\{\tau\},\cD,
	    \mathsf{check}(\tau,s)\cons h,\tuple{p,s[\tau]}}$.
	 Therefore, we have $rev(\tuplef{\cC,\cD,h,\tuple{p,s[\mathsf{check}]}}) =
	 rev(\tuplef{\cC\cup\{\tau\},\cD,
	    \mathsf{check}(\tau,s)\cons h,\tuple{p,s[\tau]}})
	    = \tuplef{h,\tuple{p,s[ok]}}$.
	    Observe that the item $\mathsf{check}(\tau,s)$ is
	    removed from the history by function $rev$.

	\item The case for the \underline{\emph{Commit}} rules
	is perfectly analogous.
	
	\item Finally, the case of rules \underline{\emph{Delay}} and 
	\underline{\emph{Rollback}} is trivial
	since the history $h$ and the standard component $\tuple{p,s}$
	is the same before and after the application of these rules.	
	\end{itemize}
	Consider now that the selected process of $S$ has the form 
	$\tupleb{\cC,\cD,h,\tuple{p,s}}^\tau$. Here, we consider the
	following cases:
	\begin{itemize}
	\item If the applied rule is $\overline{\mathit{Seq}}$, then
	the claim follows straightforwardly since the rule is a copy of the
	corresponding rule in the reversible semantics if we omit
	arguments $\cC$ and $\cD$.
	
	\item The first rules of $\overline{\mathit{Send}}$,
	$\overline{\mathit{Receive}}$ and $\overline{\mathit{Spawn}}$
	are essentially equivalent to their counterpart in the reversible
	semantics (i.e., if we ignore the sets 
	$\cC$ and $\cD$ that are filtered away by function $rev$). 
	Therefore, the claim follows trivially.

   \item As for the second rules of $\overline{\mathit{Send}}$,
	$\overline{\mathit{Receive}}$ and $\overline{\mathit{Spawn}}$,
	either nothing is changed in the configurations (the case of
	$\overline{\mathit{Send}}$ and $\overline{\mathit{Spawn}}$)
	or we can proceed exactly as above (the case of 
	$\overline{\mathit{Receive}}$).

	\item If the applied rule is one of the $\overline{\mathit{Check}}$
	rules, we have a backward step which is either of the form\\[-1ex]
	
	$
	\tupleb{\cC,\cD,\mathsf{check}(\tau,s)\cons h,\tuple{p,s[\tau]}}^\tau
	\hoo 
	\tuplef{\cC\setminus\{\tau\},\cD,h,\tuple{p,s[\mathsf{check}]}}
	$\\[-1ex]

	or\\[-1ex]
	
	$
	\tupleb{\cC,\cD,\mathsf{check}(\tau',s)\cons h,\tuple{p,s[\tau']}}^\tau
	\hoo 
	\tupleb{\cC\setminus\{\tau'\},\cD,h,\tuple{p,s[\mathsf{check}]}}^\tau
	$\\[-1ex]

	In all four configurations, function $rev$ returns
	$\tuplef{h,\tuple{p,s[ok]}}$, and the claim follows.

	Finally, the case of the 
	application of rule $\overline{\mathit{Commit}}$
	is perfectly analogous to the previous case.

	\qed
	\end{itemize}
\end{proof}
The following lemma shows that every computation
between a checkpoint and the corresponding rollback is indeed
reversible, as expected. It will be also useful to prove some partial
completeness for the rollback semantics.

\begin{lemma} \label{lemma:history}
	Let $d = (S_0 \hoo S_1 \hoo \ldots \hoo S_n)$ be a well-defined
	derivation under the rollback semantics with 
	$S_i \hoo_{p,\underline{\mathsf{rollback}}(\tau)} S_{i+1}$, $i\leq n$.
	Then, there exists $S_j \hoo_{p,\underline{\mathsf{check}}(\tau)} S_{j+1}$,
	$j<i$, such that the set of active checkpoints of $p$
	is never empty between $S_{j+1}$ and $S_i$.
\end{lemma}

\begin{proof}
	Since the derivation is well-defined, we have that the
	step $S_j \hoo_{p,\underline{\mathsf{check}}(\tau)} S_{j+1}$,
	$j<i$, indeed exists. Moreover, by condition (4),
	we know that there cannot be an occurrence of
	$\mathsf{commit}(\tau,s)$ in the history of process
	$p$ in system $S_i$. Therefore, the set of active
	checkpoints of process $p$ between $S_{j+1}$ and $S_i$
    must contain at least $\tau$, so they are never empty.
\end{proof}
We leave the study of other interesting 
results of our rollback semantics 
(e.g., minimality and some partial completeness) for future work.

\end{document}